\documentclass[final,5p,times,twocolumn]{elsarticle}
\usepackage{amsmath, amssymb}
\graphicspath{{./figs/}}
\usepackage{caption}
\begin{document}
\begin{frontmatter}
\title{Ghost Dark Energy in Tsallis and Barrow Cosmology}

\author{Esmaeil Ebrahimi$^{1,2,3}$\corref{mycorrespondingauthor}}
\ead{es.ebrahimi@saadi.shirazu.ac.ir}

\author{Ahmad Sheykhi$^{1,2}$}
\ead{asheykhi@shirazu.ac.ir}

\address{$^{1}$Department of Physics, College of  Science, Shiraz University, Shiraz 71454, Iran.}
\address{$^{2}$Biruni Observatory, College of Science, Shiraz University, Shiraz 71454, Iran.}
\address{$^{3}$College of Physics, Shahid Bahonar University, PO Box 76175, Kerman, Iran}
\cortext[mycorrespondingauthor]{Corresponding author}


\begin{abstract}
According to the thermodynamics-gravity conjecture, any
modification to the entropy expression leads to the modified
cosmological field equations. Based on this, we investigate the
cosmological consequences of the modified Friedmann equations when
the entropy associated with the horizon is in the form of
Tsallis/Barrow entropy and the dark energy (DE) component is in
the form of ghost dark energy (GDE). We perform a dynamical system
analysis and see that the Tsallis GDE(TGDE) and Barrow GDE(BGDE)
can exhibit a correct phase space evolution for suitable range of
the free parameters($0<\Delta<1$ for BGDE and $\beta<3/2$ for
TGDE). We find that in BGDE and TGDE (with $Q=3b^2
H(\rho_D+\rho_m)$), there exist an early radiation dominated phase
of expansion which is absent for GDE model in standard cosmology.
It is worth mentioning that seeking an unstable phase of matter
dominated in TGDE leads to a new constraint on $\beta$($<3/2$).
Using the resulting range of free parameters from latter step we
find that both models are capable to explain the cosmic evolution
from deceleration to an accelerated phase. We observe that
increasing $\Delta$ ($\beta$) parameters leads to a delay in the
cosmic phase transition. We conclude that BGDE and TGDE are viable
cosmological models which predict a consistent phase transition of
the cosmic expansion for suitable ranges of the parameters. We
also calculate the squared sound speed for both models and find
out that they are unstable against perturbations. Next, we proceed
the statefinder analysis and see that both models are
significantly distinct from $\Lambda CDM$ as well as with respect
to each other at past epochs, while both of them catch the
standard cosmology at far future. We also explore the impacts of
the non-extensive parameters on the density perturbations and find
out that the pattern of density contrast ($\delta$) evolution in
TGDE/BGDE is distinct from GDE in standard cosmology.
\end{abstract}

\begin{keyword}
Modified cosmology\sep Ghost dark energy\sep Tsallis entropy\sep
Barrow entropy
\end{keyword}

\end{frontmatter}

\section{Introduction}\label{intro}
Modified gravity is an alternative explanation for the late time
acceleration of the cosmic expansion
\cite{riess982,riess98,SupernovaCosmologyProject:1998vns,WMAP:2003elm}.
Modified theories of gravity including generalization of the
Einstein gravity such as $f(R)$
\cite{Capozziello:2002rd,Sobouti:2006rd,Sotiriou:2008rp,Boehmer:2007kx,
Cognola:2007zu,Nojiri:2010wj,Capozziello:2011et,Odintsov:2019evb,Odintsov:2019mlf},
the Gauss-Bonet gravity
\cite{Lanczos1932ElektromagnetismusAN,Lanczos1938ARP,Dehghani:2004cf}
and Lovelock theory of gravitation
\cite{Lovelock1969DivergencefreeTC}. The main idea is to explain
the early inflation epoch as well as the late time acceleration of
the cosmic expansion via modified gravity.

On the other hand, cosmological models based on the modified
entropy expression have got a lot of attentions in recent years.
According to thermodynamics of black holes, the entropy and
temperature associated with the horizon are proportional to the
horizon area and surface gravity, respectively. Is there any
direct connection between the gravitational field equations and
the laws of thermodynamics? Jacobsen was the first who answered
this question and derived the Einstein's field equations of
gravity from Clausius relation $\delta Q=T \delta S$, where
$\delta Q$ is the heat flux crossing the horizon, $T$ and $S$ are,
respectively, temperature and entropy of the horizon
\cite{Jacobson:1995ab}. The next great step put forwarded by Eric
Verlinde \cite{Verlinde:2010hp} who claimed that gravity is not a
fundamental force and can be regarded as an entropic force. In
this approach gravitational interaction originates from changes of
the entropy associated with the information on the holographic
screen. Following \cite{Verlinde:2010hp} many efforts have been
done to explore the entropic nature of gravity
\cite{Cai:2010hk,Cai:2010sz,Banerjee:2010yd,Liu:2010na,Sheykhi:2010wm,
Modesto:2010rm,Cai:2010zw,Hendi:2010xr,Sheykhi:2010yq}.

In recent years a lot of attentions have arisen to the
non-extensive entropy formalism, where total entropy of a system
cannot necessarily obtained via summation of that of the
subsystems \cite{Tsallis:1987eu,Lyra:1997ggy,Wilk:1999dr}. The so
called Tsallis entropy is one of the most well-known case of the
non-extensive entropies. In 2012, Citro and Tsallis discussed that
entropy associated with the horizon of a black hole does not obey
the area law and instead may be written as \cite{Tsallis:2012js},
\begin{equation}
S_h=\gamma A^{\beta}
\end{equation}
where $A$ is the black hole horizon area, $\gamma$ is an unknown
constant and $\beta$ is known as Tsallis (or non-extensive)
parameter. We assume $\beta$ is a real parameter which shows the
amount of the non-extensivity \cite{Tsallis:2012js}. It can be
easily seen that the usual area law of black hole entropy is
deduced in the limiting case where $\beta\rightarrow1$ and
$\gamma\rightarrow1/(4l_p^2)$.

Regarding this new form of entropy and its cosmological
consequences, two approaches has been followed in the literature.
The first approach only modifies the DE density due to the change
in the entropy, while keeping the cosmological field equations in
its standard form. The second approach modifies the cosmological
field equations through modified entropy, beside to plausible
corrections to the DE density. Since the holographic dark energy
(HDE) is based on the definition of black hole entropy, thus
redefinition of the entropy will affect the resulting HDE. In this
regard the modified HDE in the context of standard cosmology has
been extensively explored
\cite{Tavayef:2018xwx,Zadeh:2018poj,AbdollahiZadeh:2019lsx,Pandey:2021fvr}.
It was shown that for a special choice of the parameter $\beta$
the Tsallis HDE leads to an accelerated universe and can alleviate
the age problem \cite{Huang:2019hex}. On the other side, the
thermodynamic-gravity conjecture reveals modifications in the
resulting Friedmann equations. Modified Friedmann equations
through Tsallis entropy were explored in \cite{Sheykhi:2018dpn}.
Different aspects of Tsallis cosmology were investigated and the
model were also constrained with observational data
\cite{Asghari:2021lzu}. Growth of perturbations in Tsallis
cosmology was explored in \cite{Sheykhi:2022gzb}. In
\cite{Basilakos:2023kvk}, the authors have argued that Tsallis
cosmology can alleviate $H_0$ and $\sigma_8$ tensions.
Baryogenesis in Tsallis cosmology is performed in
\cite{Luciano:2022ely} and the authors showed that induced effects
of non-extensivity to the Hubble function evolution in earlier
epochs can lead a mechanism to allow the baryogenesis.

Inspired by the fractal structure of the corona virus, Barrow
proposed a fractal structure for the horizon geometry of black
holes  \cite{Barrow:2020tzx}. This fractal nature of the horizon,
results to a finite volume and an infinite (or a finite) area.
Such a geometry for horizon leads to a non-extensive entropy
relation called Barrow entropy, which is given by
\begin{equation}
S=\left(\frac{A}{A_0}\right)^{1+\frac{\Delta}{2}},
\end{equation}
where $A$ and $A_0$ are the black hole horizon and Planck area,
respectively. In the above relation $\Delta$ stands for the amount
of the quantum-gravitational deformation effects and ranges
$\in[0,1]$. When $\Delta=0$, the standard area law is restored and
$A_0 \longrightarrow 4G$, while $\Delta = 1$ represents the most
intricate and fractal structure of the horizon. Modified Friedmann
equations from Barrow entropy were introduced in
\cite{Sheykhi:2021fwh,Saridakis:2020lrg}. Applying the Barrow
entropy, validity of the generalized second law of thermodynamics
has argued in \cite{Saridakis:2020cqq}. It is worth mentioning
that Barrow cosmology cannot lead to a consistent picture of
cosmic evolution in the absence of the DE component. Thus
exploring Barrow cosmology with DE component is well motivated.
For example, HDE in Barrow cosmology as well as observational
constraints on these models have been investigated
\cite{Saridakis:2020zol,Srivastava:2020cyk,Adhikary:2021xym,Oliveros:2022biu,Anagnostopoulos:2020ctz,Dabrowski:2020atl}.
Structure formation in the context of Barrow cosmology was
addressed in \cite{Sheykhi:2022gzb}.

It is important to note that although Barrow and Tsallis relations
for the entropy expressions are almost similar but they arise from
completely different perspective. Tsallis entropy originates from
generalizing standard thermodynamics, while Barrow entropy
correction to the entropy comes from quantum gravitational
effects. It is also worth noting that $\Delta$ and $\beta$
parameters, which denote the degree of non-extensivity, lie in
different ranges ($\Delta \epsilon$ [0,1] and $\beta
\epsilon$[0,2]) for each model.

On the other side, the GDE model was proposed to alleviate the QCD
U(1) problem
\cite{veneziano1979u,Witten:1979vv,Rosenzweig:1979ay,Nath:1979ik,Kawarabayashi:1980dp}.
Vacuum energy contribution of QCD field can act as a DE component
in the curved spacetime or in a dynamic background. Up to first
order the resulting vacuum energy in an expanding universe leads
to $\rho_D\propto \Lambda_{QCD}^3 H$, where $H$ is the Hubble
expansion rate and $\Lambda_{\rm QCD}^3$ is QCD mass scale
\cite{Ohta:2010in}. Taking $\Lambda_{QCD} \sim 100MeV$ and $H_0
\sim 10^{-33}eV$, the resulting vacuum energy density equals to
$\rho_{GDE}\sim 10^{-3}eV$ in favor of observations.

It is important to note that in GDE model, we do not need
to introduce any new degree of freedom and thus this model is
completely preferred with respect to those add new degrees of
freedom. First interesting out coming of this model is its 'fine
tuning'-free nature \cite{Urban:2009vy,Ohta:2010in}. Different
aspects of GDE have been investigated in standard cosmology. While
dynamics of the universe in presence of the GDE and DM was studied
in \cite{Cai:2010uf}, interacting GDE in a curved background was
discussed in \cite{Sheykhi:2011xz}. Stability of a GDE dominated
universe was addressed in \cite{Urban:2009vy,Ebrahimi:2011js}.
Also a statefinder analysis of GDE model explored in
\cite{Malekjani:2012wc}. Correspondence between the interacting
GDE and a complex quintessence cosmology was performed in
\cite{Liu:2020bmp}.

Although taking a non-extensive form of the entropy in Barrow and
Tsallis cases modifies the Friedmann equations, however the
resulting Barrow and Tsallis cosmology are not well satisfying
specially in facing the accelerated expansion of the universe.
It is worth mentioning that in \cite{Sheykhi:2018dpn} one
of the present authors argued that for $\beta<1/2$, Tsallis
cosmology can explain the late time accelerated expansion without
invoking any kind of dark energy. Nevertheless, in
\cite{Asghari:2021lzu}, the authors found an observational
constraint on Tsallis parameter($\beta$) and the best fit value
for Tsallis parameter is ($\beta\approx 0.999$) which denotes a
discrepancy between Tsallis cosmology (in the absence of any DE
component) and observational evidences. Thus considering an
additional component of energy (such as GDE) in these modified
cosmological models, is still well motivated. From a closer look
one should note that two important feature in DE models are still
of main motivations in exploring the GDE in Barrow and Tsallis
cosmology. First is that the Barrow (Tsallis) HDE as well as
Barrow ADE suffers instability issues regarding the squared sound
speed analysis
\cite{Zadeh:2018poj,Sheykhi:2022fus,Sheykhi:2023woy}. Second
reason comes from the behavior of the models in facing the phantom
line. For example the Barrow ADE can not cross this
border\cite{Zadeh:2018poj,Sheykhi:2022fus,Sheykhi:2023woy}. Hence,
there exist enough motivations to describe the GDE in Barrow and
Tsallis cosmology. Thus, we would like to explore the influences
of the entropy based modified cosmological models when the DE
component of the universe is taken as GDE. We shall explore the
cosmological consequences of this model in the context of modified
Barrow/Tsallis modified cosmology and study the evolution of the
cosmological parameters.

The outlook of the article is as follows. In the next section,
using the thermodynamics-gravity correspondence, we review
derivation of the modified Friedmann equations based on
Tsallis/Barrow entropy. In section \ref{Barrow}, we investigate
GDE in the context of the modified Barrow cosmology. In section
\ref{Tsallis}, we explore GDE in the context of the modified
Tsallis cosmology. The last section is devoted to closing remarks.

\section{Tsallis and Barrow cosmology}
In this section, we derive the modified Friedmann equations using
the first law of thermodynamics at apparent horizon of a FRW
universe. To this goal we assume that the entropy associated with
the apparent horizon is in the form of Barrow/Tsallis entropy. We
take the spatial subspace of the universe as a maximally symmetric
space. In this case the form of the line element reads
\begin{equation}\label{ds}
ds^2=h_{\mu\nu}dx^{\mu}dx^{\nu}+\tilde{r}^2(d\theta^2+sin^2\theta
d\phi^2),
\end{equation}
where $\tilde{r}=a(t)r, x^0=t, x^1=r$ and $h_{\mu\nu}=
diag\left(-1,\frac{a^2(t)}{1-kr^2}\right)$ is two dimensional
metric. $k=-1,0,+1$ corresponds to open, flat and closed universe
respectively. In order to choose a boundary at which first and
second laws of thermodynamics hold on, the suitable choice is
apparent horizon. The definition of the apparent horizon is
\cite{Cai:2009ph}
\begin{equation}\label{app}
\tilde{r}_{A}=\frac{1}{\sqrt{H^2+{k}/{a^2}}}.
\end{equation}
where $H=\dot{a}/a$ is the Hubble parameter. The resulting
temperature with the apparent horizon is \cite{Cai:2009ph}
\begin{equation}\label{temp}
T_h=-\frac{1}{2\pi\tilde{r}_A}\left(1-\frac{\dot{\tilde{r}}_A}{2H\tilde{r}_A}\right).
\end{equation}
Further, we assume the cosmic fluid could be described through
\begin{equation}\label{tmn}
T_{\mu\nu}=(\rho+p)u_{\mu}u_{\nu}+pg_{\mu\nu},
\end{equation}
where $\rho$ and $p$ are the energy density and pressure of the cosmic fluid respectively.
Conservation equation $(\nabla_{\mu}T^{\mu\nu}=0)$ leads to the continuity equation as
\begin{equation}\label{coneq}
\dot{\rho}+3H(\rho+p)=0.
\end{equation}
In an expanding background any change in the volume element is
associated with the following amount of the work density
\cite{Hayward:1997jp}
\begin{equation}\label{work}
W=-\frac{1}{2}T^{\mu\nu}h_{\mu\nu}.
\end{equation}
Above definition of work density in terms of the energy momentum tensor elements, (\ref{tmn}), becomes
\begin{equation}\label{wdef}
W=\frac{1}{2}(\rho-p).
\end{equation}
Now we are ready to extract Friedman equations through first law
of thermodynamics. On this way, we start with the following
equation
\begin{equation}\label{flaw}
dE=T_hdS_h+WdV,
\end{equation}
and assume that above equation is satisfied on the apparent
horizon(\ref{app}). Here $V=4\pi\tilde{r}^3/3$ is the volume of a
sphere with radius $\tilde{r}$, $S_h$ is the associated entropy
with apparent horizon and $E=\rho V$ is the total energy of the
universe.  Differentiating $E=\rho V$ and mixing the result with
the continuity equation ($d\rho=-3H(\rho+p)dt$), on gets
\begin{equation}\label{der}
dE=4\pi \tilde{r}^2 \rho d\tilde{r}-4\pi H \tilde{r}^3(\rho+p)dt
\end{equation}
Given this equation at hand, one can proceed to get the evolution
equation of any expanding universe with an arbitrary entropy
definition. In this paper we would like to obtain the resulting
equation for Barrow/Tsallis entropy definitions.

The Barrow entropy formula reads \cite{Barrow:2020tzx}
\begin{equation}\label{barrowen}
S=\left(\frac{A}{A_0}\right)^{1+\frac{\Delta}{2}}.
\end{equation}
Taking differential form of this equation and using Eqs.
(\ref{temp}), (\ref{flaw}) and (\ref{der}), after a little algebra
one arrives at
\begin{equation}\label{bardef}
-\frac{2+\Delta}{2\pi
A_0}\left(\frac{4\pi}{A_0}\right)^{\frac{\Delta}{2}}\frac{d\tilde{r}}{\tilde{r}^{3-\Delta}}=\frac{d\rho}{3}.
\end{equation}
Integrating the above equation and using (\ref{app}), we reach
\cite{Sheykhi:2021fwh}
\begin{equation}\label{barfried}
\left(H^2+\frac{k}{a^2}\right)^{1-\Delta/2}=\frac{2\pi
A_0}{3}\left(\frac{2-\Delta}{2+\Delta}\right)\left(\frac{A_0}{3}\right)^{\Delta/2}\rho,
\end{equation}
which can be rewritten as \cite{Sheykhi:2021fwh}
\begin{equation}\label{barfriedf}
\left(H^2+\frac{k}{a^2}\right)^{1-\Delta/2}=\frac{8\pi G_{\rm
eff}}{3}\rho,
\end{equation}
where the effective gravitational constant $G_{\rm eff}$ is
defined as \cite{Sheykhi:2021fwh}
\begin{equation}\label{geff}
G_{\rm
eff}=\frac{A_0}{4}\left(\frac{2-\Delta}{2+\Delta}\right)\left(\frac{A_0}{3}\right)^{\Delta/2}.
\end{equation}
It could be easily seen that in the limit $\Delta\rightarrow0$ and
$A_0\rightarrow 4G$, we have $G_{\rm eff}\rightarrow G$, and the
standard Friedmann equation is retrieved.

Starting with definition of the Tsallis entropy relation,
$S=\gamma A^{\beta}$, and performing an almost same set of
derivations the modified Friedmann equation in Tsallis case can be
obtained. The resulting first Friedmann equation in Tsallis case
reads \cite{Sheykhi:2018dpn}
\begin{equation}\label{tfriedf}
\left(H^2+\frac{k}{a^2}\right)^{2-\beta}=\frac{8\pi
l^{2}_{P}}{3}\rho,
\end{equation}
where $l^2_{p}=\frac{2-\beta}{4\beta\gamma}(4\pi)^{1-\beta}$. One
should note that in the limiting case $\beta=1$ and
$\gamma=1/(4l^{2}_{p})$ the standard Friedmann equation is
recovered. Since in present work we do not use the second
Friedmann equation we do not mention about that. However a
detailed derivation of the Tsallis and Barrow cosmology equations
can be seen in
\cite{Sheykhi:2018dpn,Sheykhi:2021fwh,Saridakis:2020lrg}.

It is worthy to note that despite the non-extensive
statistics modifications in the resulting cosmology, the new
frameworks are still in need of an additional DE component to
overcome the late time acceleration issue conveniently
\cite{Sheykhi:2021fwh}. Thus it is still well motivated to
consider the GDE in the context of Tsallis/Barrow modified
cosmology.
\section{Ghost dark energy in Barrow cosmology} \label{Barrow}
Here, we would like to consider a flat universe filled with a
pressureless matter and the GDE components. In this case the first
Friedmann equation in the modified Barrow cosmology reads
\begin{align}\label{frideq1}
H^{2-\Delta}=\frac{8\pi G_{\rm eff}}{3}(\rho_m+\rho_D),
\end{align}
where $\rho_m$ and $\rho_D$ are energy density of matter and GDE,
respectively. In an expanding background the continuity equations
are
\begin{align}\label{conseq}
\dot{\rho}_{D}+3H\rho_{D}(1+w_{D})&=-Q ,\\
\dot{\rho}_{m}+3H\rho_{m}&=Q,
\end{align}
where $w_{D}=p_{D}/\rho_{D}$ is equation of state parameter (EoS)
of dark component and $Q$ denotes the chance of interaction
between matter and DE component. One should note that the
DE and DM components are introduced to literature for curing
different issues and these component are both separately
conserved. However, in the absence of a symmetry regarding
separate conservation, people give a chance to mutual interaction
between DE and DM. Further, presence of interaction between DE and
DM makes a closer agreement to observational evidences from the
galaxy cluster Abell A586\cite{Bertolami:2007zm}. Thus according
to these theoretical and observational reasons it is well
motivated to consider an interaction between dark sector
components. Interacting models were first presented by Wetterich
in\cite{Wetterich:1987fm}. Simplest form of interaction term has
the form $Q\propto H \rho$, where $\rho$ could be $\rho_D$,
$\rho_m$ or $\rho_m+\rho_D$. Various models of dark energy are
explored in presence of such interaction
terms\cite{Amendola:1999qq,Amendola:2000uh,Zimdahl:2001ar,
Zimdahl:2002zb,Chimento:2003iea,Wang:2005jx,Wang:2005ph}.
Here, we will consider the choice $Q=3b^2H(\rho_D+\rho_m)$ in
Eq.(\ref{conseq}). Beside it worth to mention that some non-linear
choices of $Q$ has been also examined in the
literature\cite{Mangano:2002gg,Arevalo:2011hh,Baldi:2010vv}.

The fractional energy densities could be defined as
\begin{equation}\label{Omega}
\Omega_m=\frac{\rho_m}{\rho_{cr}},\ \ \
\Omega_D=\frac{\rho_D}{\rho_{cr}},
\end{equation}
where the critical energy density is
$\rho_{cr}=3H^{2-\Delta}/(8\pi G)$. Thus, the first Friedmann
equation in terms of the fractional energy densities reads
\begin{equation}\label{fridomega}
\Omega_m+\Omega_D=1.
\end{equation}

Our main mission in this paper is to disclose the impacts of the
GDE in entropy corrected cosmology frameworks. As we mentioned in
the introduction, the energy density of the GDE in an expanding
background is written as \cite{Cai:2009ph}
\begin{equation}\label{gdef}
\rho_D=\alpha H,
\end{equation}
where $\alpha$ denotes a constant of order $\Lambda_{QCD}^3$. At
first step we consider the EoS parameter for the GDE in the
modified Barrow cosmology. Solving the continuity equation for
$w_D$, one finds
\begin{equation}\label{wd1}
w_D=-1-\frac{\dot{\rho}_D}{3H\rho_D}-\frac{Q}{3H\rho_D}.
\end{equation}

Taking the time derivative from Eqs. (\ref{gdef}) and
(\ref{frideq1}) and using the fractional density definitions, one
gets
\begin{equation}\label{brhobrrhodot}
\frac{\dot{\rho}_D}{\rho_D}=\frac{\dot{H}}{H}=-\frac{3H(\Omega_Dw_D+1)}{2-\Delta}.
\end{equation}
Taking Eq. (\ref{wd1}) into account and using relation $Q=3b^2
H(\rho_D+\rho_m)$, one finds
\begin{equation}\label{bwdi}
w_D=-\frac{1}{2-\Omega_D-\Delta}\left[1-\Delta+\frac{b^2}{\Omega_D}(2-\Delta)\right].
\end{equation}
One can easily check that for $\Delta=0$, the above relation for
$w_D$ reduces to the corresponding relation in GDE
\cite{Cai:2010uf}. The deceleration parameter is another important
quantity in our setup. Using definition of the deceleration
parameter as
\begin{equation}\label{qdef}
q=-1-\frac{\dot{H}}{H^2}=-1+\frac{3}{2-\Delta}\left(1+\Omega_D
w_D\right),
\end{equation}
and substitute Eq. (\ref{bwdi}), we get
\begin{equation}
q=-1+\frac{3(b^2+\Omega_D-1)}{\Omega_D-2+\Delta}
\end{equation}
Evolution of the DE fractional density can complete our equation
set. To this end we use $\dot{\Omega}_D=H \Omega^{'}_D$ which
prime denotes a differentiation with respect to $x=\ln{a}(a_0=1)$.
Taking a time derivative of $\Omega_D$ one can get
\begin{equation}\label{omegaev}
\Omega^{'}_D=(1-\Delta)\frac{3\Omega_D(b^2+\Omega_D-1)}{\Omega_D-2+\Delta}
\end{equation}
Once again one can check that above equations approach a correct
limit for $\Delta=0$ in both interacting and non-interacting cases
in agreement with \cite{Cai:2010uf}.

\subsection{Dynamical system analysis of Barrow cosmology}
A primary description in disclosing the BGDE cosmology is the
phase-space analysis. This approach is widely discussed in
cosmological and astrophysical
frameworks\cite{Amendola:1999er,xu2012phase,Landim:2015uda,Landim:2015poa}.
The dynamical system analysis allows us to obtain a qualitative
description of dynamics of cosmological models. This analysis
could be performed in ample details but our aim here is to see
overall features of the model. Specifically, we would like to see
if the BGDE is capable to result in radiation, DM and DE dominated
phases with correct timeline order. Beside, we can obtain a prior
range of free parameters through this technique. On this way, we
take the universe filled by radiation, CDM and the GDE. Next, we
introduce \{$\Omega_D,\Omega_m$\} as the dynamical variables and
obtain the corresponding evolution equations
\{$\Omega^{\prime}_D=0,\Omega^{\prime}_m=0$\}. Then algebraic
solution of the autonomous set of the first order equations
reveals evolutionary behavior of each
model\cite{Amendola:1999er,xu2012phase,Landim:2015uda,Landim:2015poa,Leon:2021wyx}.

We already introduced the evolution equation of $\Omega_D$ in
Eq.(\ref{omegaev}). Performing a same calculations for $\Omega_m$,
we will obtain the resulting dynamical autonomous system of the
BGDE as
\begin{equation}\label{omdprimeb}
\Omega^{\prime}_D=\frac{\Omega_{\mathrm{D}} \left(3 b^{2} \Omega_{\mathrm{D}}+3 b^{2} \Omega_{m}+4 \Omega_{\mathrm{D}}+\Omega_{m}-4\right) \left(\Delta -1\right)}{\Omega_{\mathrm{D}}-2+\Delta},
\end{equation}

\begin{equation} \label{ommprimeb}
\Omega^{\prime}_m=\frac{\left(f_{1b} \Omega_{m}^{2}+f_{2b} \Omega_{m}+3 b^{2} \Omega_{\mathrm{D}} \left(\Omega_{\mathrm{D}}-2+\Delta \right)\right)}{\Omega_{\mathrm{D}}-2+\Delta},
 \end{equation}
 where
 \begin{eqnarray}
 f_{1mb}&=&\left(-3 \Delta +6\right) b^{2}+2-\Delta, \nonumber\\
 f_{2mb}&=&\left(\left(-3 \Delta +9\right) \Omega_{\mathrm{D}}+3 \Delta -6\right) b^{2}\nonumber\\&+&\left(-4 \Delta +5\right) \Omega_{\mathrm{D}}+\Delta -2.
 \end{eqnarray}
 \begin{table}
\begin{tabular}{|c|c|c|c|c|c|}
  \hline
  fixed point & $\Omega_D$ & $\Omega_m$ & $q$ & Nature \\
  \hline\hline
  $P_1$ & 0 & 0 & $>0$ & unstable for $0<\Delta<1$  \\\hline
  $P_2$ & 0 & 1 & $>0$ & saddle $0<\Delta<1$\\\hline
  $P_3$ & $-b^2+1$ & $b^2$& $<0$ & stable \\
  \hline
  \end{tabular}
  \caption{The admitted fixed points of BGDE in presence of an interaction term($Q=3b^2H(\rho_D+\rho_m)$).}
  \label{Binttab}
\end{table}

The resulting fixed points are listed in Tab.(\ref{Binttab}). The physical phase space is compact(i.e. $0\leq \Omega_i \leq 1$).
The first point($P_1$) denotes a radiation dominated phase at the
beginning of the cosmic evolution. It is worth to mention that in
\cite{Golchin:2016yci} the interacting GDE(with
$Q=3b^2H(\rho_D+\rho_m)$) in standard cosmology does not admit a
radiation dominated epoch while in Barrow cosmology this point is
present. For this fixed point the corresponding eigenvalues(whose
their signs determine the stability) are $\lambda_1=3b^2+1$ and
$\lambda_2=\frac{4 \left(\Delta -1\right)}{\Delta -2}$. It is
evident that this fixed point represents an unstable character for
all acceptable(It is
accepted that $b^2<1$ and $0<\Delta<1$). Using the deceleration
parameter $q=-1+\frac{3(b^2-1)}{\Delta-2}$, one can easily find  that at this point the
universe shows a decelerative behavior as we expect.

Second fixed point($P_2:\Omega_m=1,\Omega_D=0$), corresponds to a
matter dominated universe. The eigenvalues of the stability matrix are ($\lambda_1=-3 b^{2}-1, \lambda_2=-3\frac{\Delta -1}{\Delta -2}(b^2-1)$). Taking $\Delta<1$ and $b^2<1$(which both are favored by literature), $P_2$ is a saddle fixed point. Thus the universe will have a transition to an other available fixed point. In this era the deceleration parameter is still positive and has a same form$q=-1+\frac{3(b^2-1)}{\Delta-2}$ as the $P_1$. At this point $w_{eff}=\Omega_D w_D \approx 0^{-}$ which corresponds to a matter dominated epoch.

\begin{figure}
\centering
\includegraphics[scale=0.35]{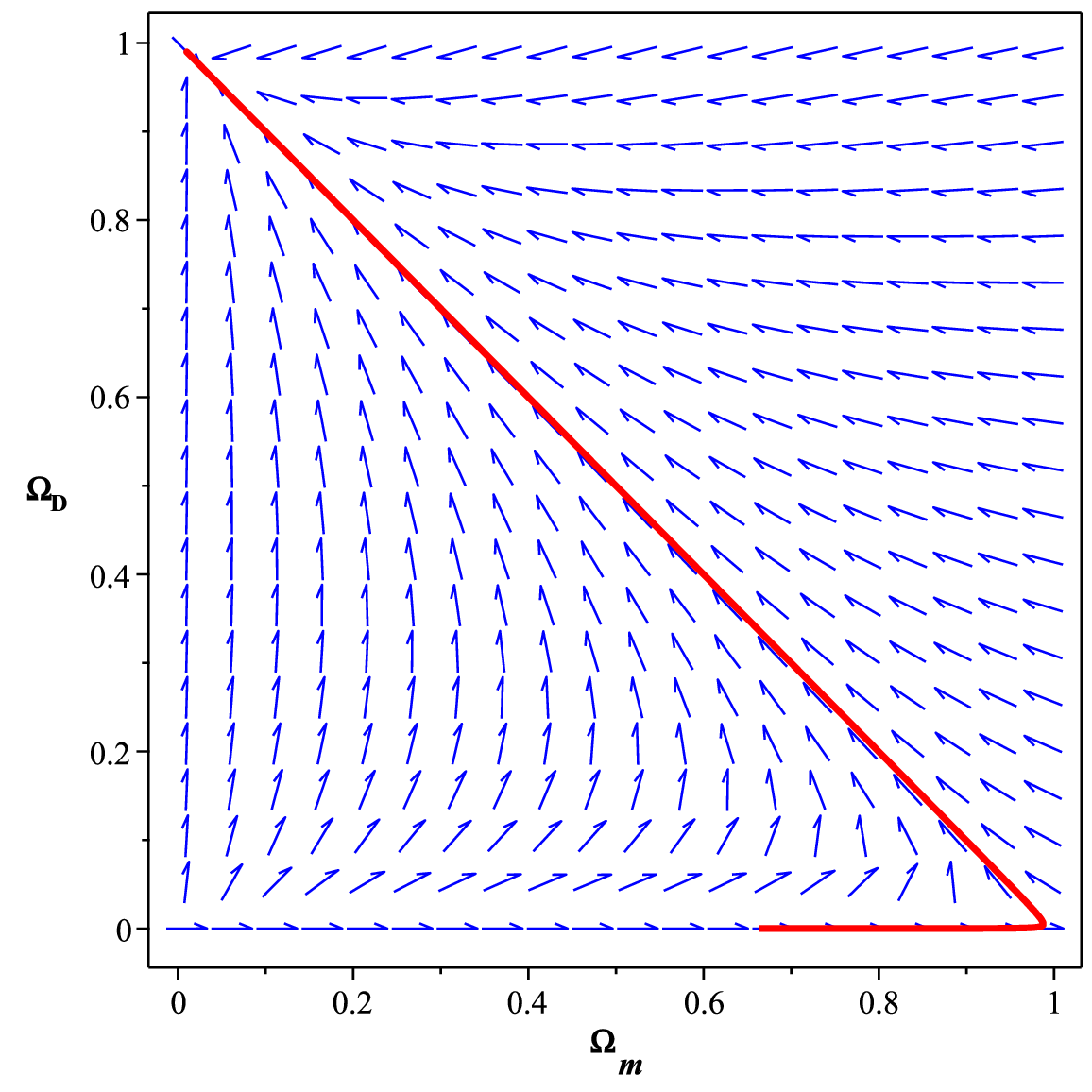}
\vspace{2mm}
\caption{Phase space evolution of the BGDE model. We set $b^2=0.01$ and take $\Delta=0.1$. The solid red line identifies the universe with $\Omega_{D0}=0.69$ and $\Omega_{m0}=0.30996$} \label{bpspace}
\end{figure}

Last fixed point($P_3: \Omega_m=b^2,\Omega_D=-b^2+1$) is evidently a late time DE dominated attractor. According to eigenvalues,$\lambda_1=-4$ and $\lambda_2=3\frac{\Delta-1}{\Delta-b^2-1}(b^2-1)$, this epoch is stable for favored ranges of free parameters $0<\Delta<1$ and $b^2<1$. In this era the universe undergoes an accelerated expansion($q=-1(w_{eff}=-1)$) irrespective of the free parameters. For completeness we plotted the phase space diagram of ($\Omega_m,\Omega_D$) in fig.(\ref{bpspace}).

\subsection{Numerical demonstration of the BGDE}
\begin{figure}
\centering
\includegraphics[scale=0.25]{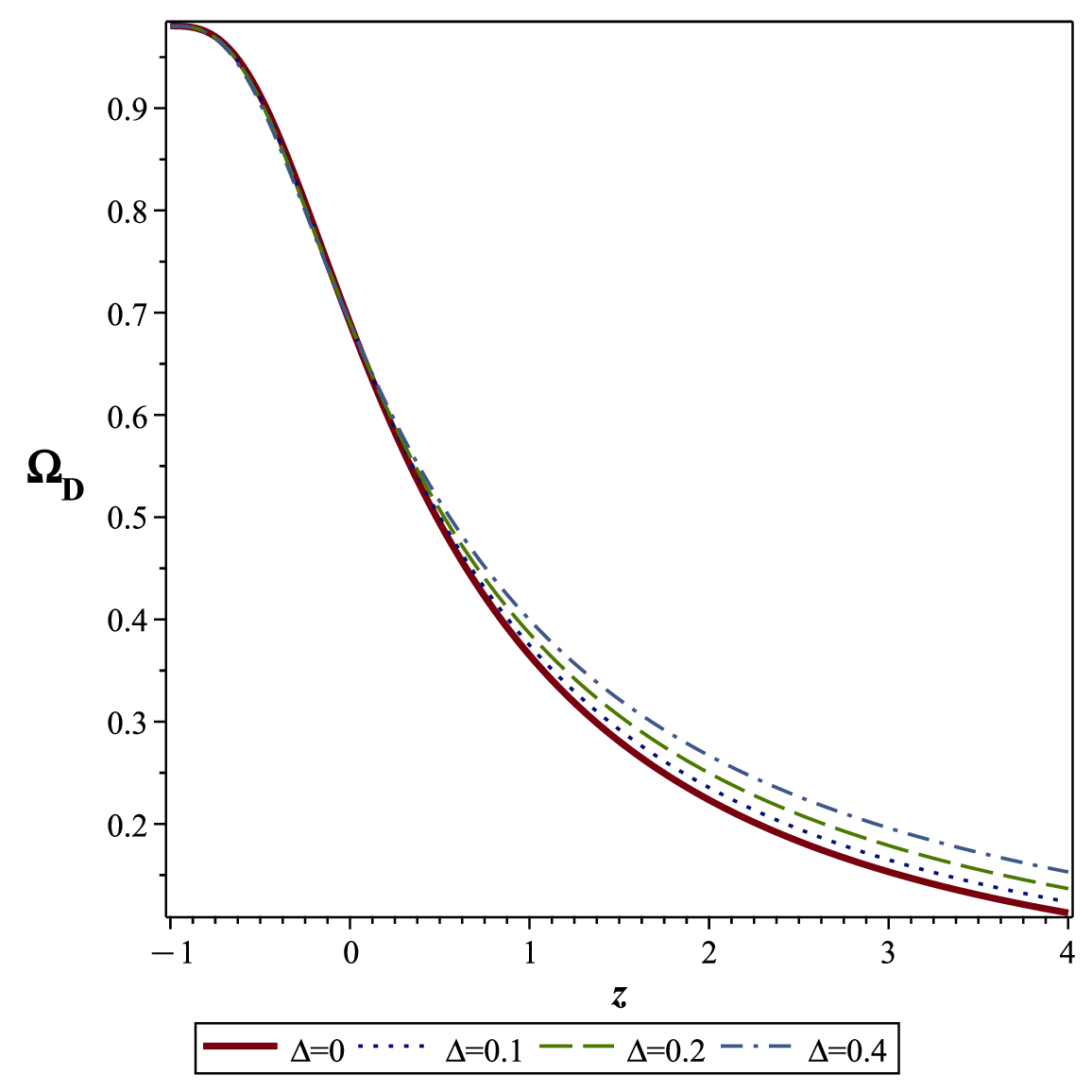}
\includegraphics[scale=0.25]{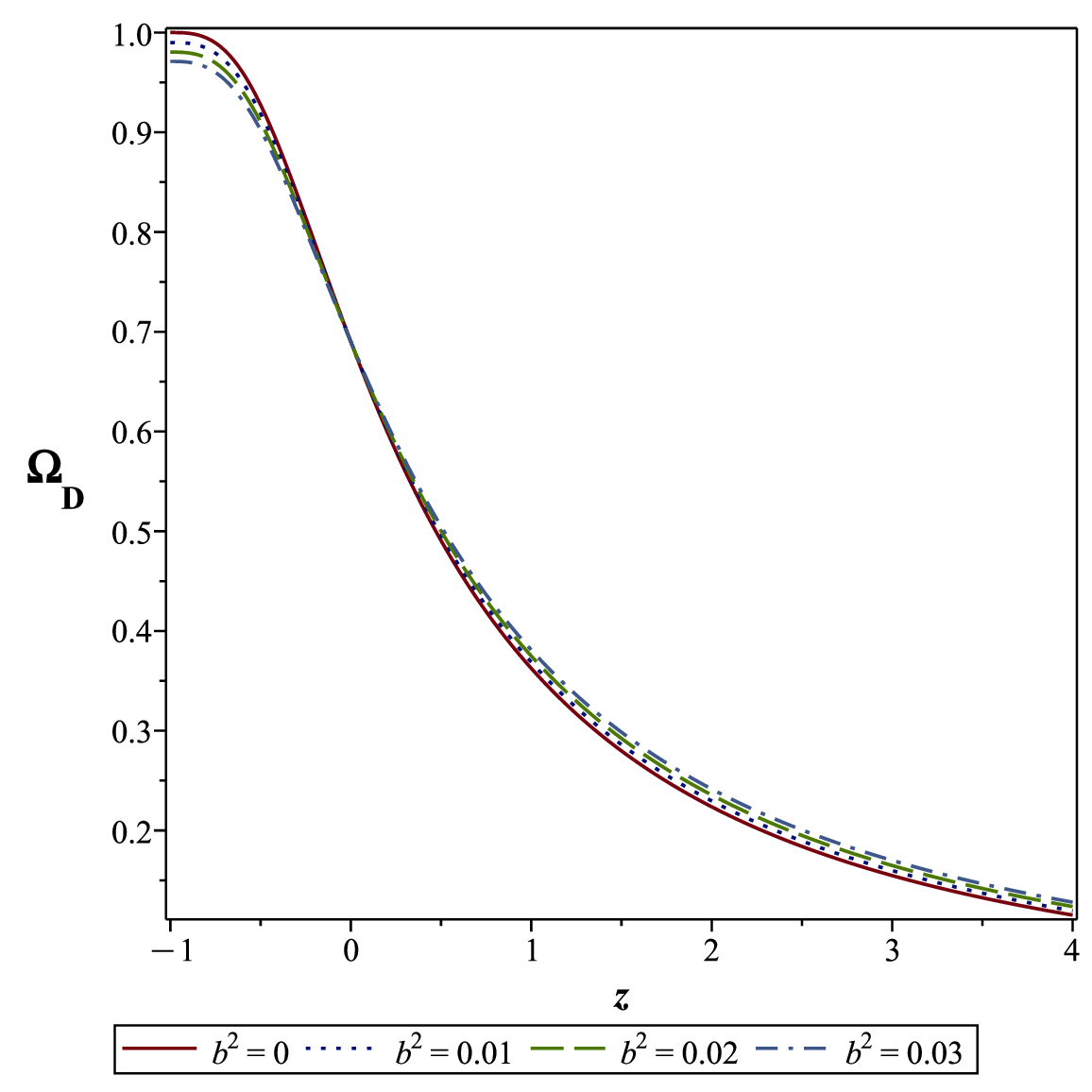}
\vspace{2mm} \caption{Evolution of $\Omega_D$ parameter for GDE in
Barrow cosmology. In upper panel we set $b^2=0.02$ while in lower
panel we take $\Delta=0.1$.} \label{omegfbdelvar}
\end{figure}

\begin{figure}
\centering
\includegraphics[scale=0.25]{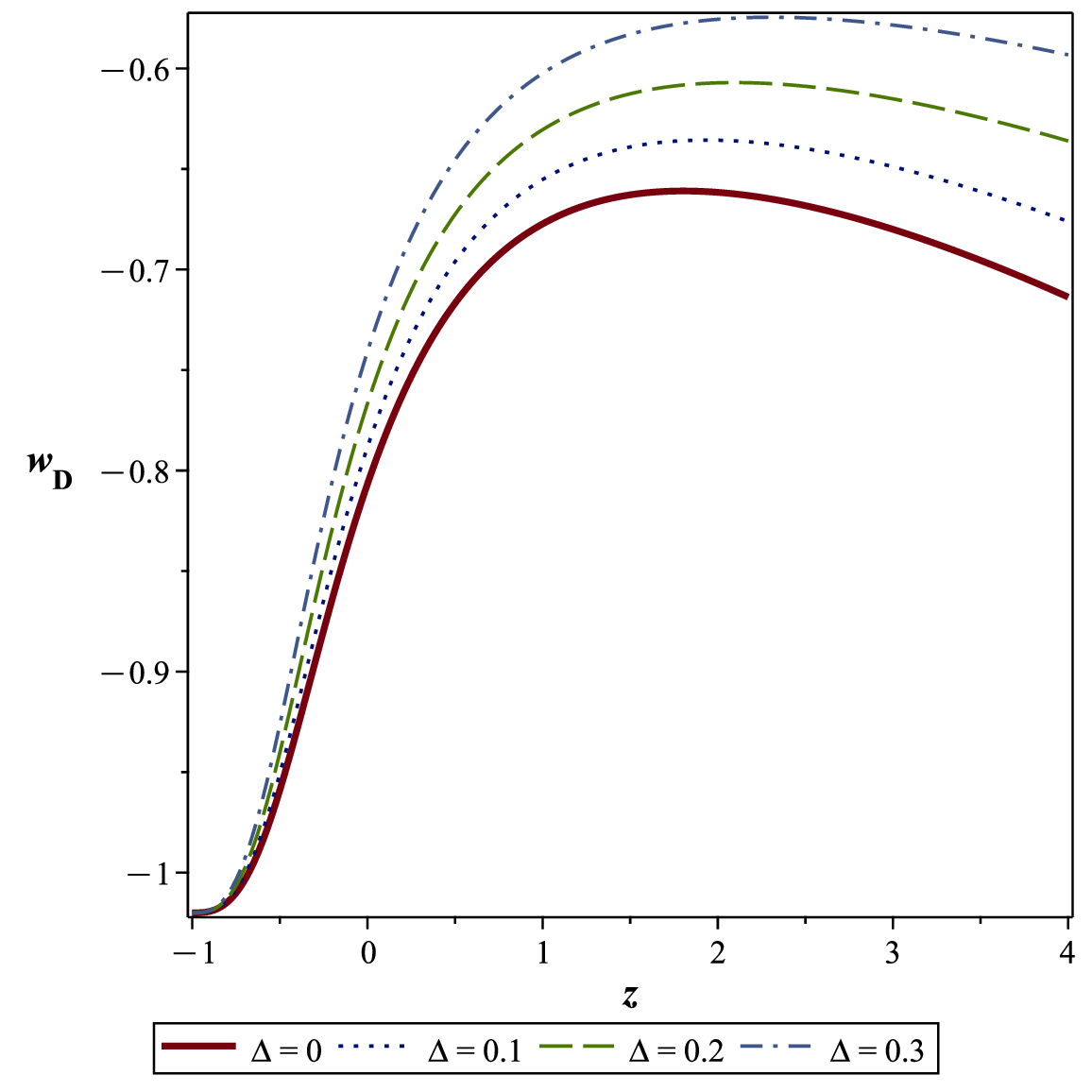}
\includegraphics[scale=0.25]{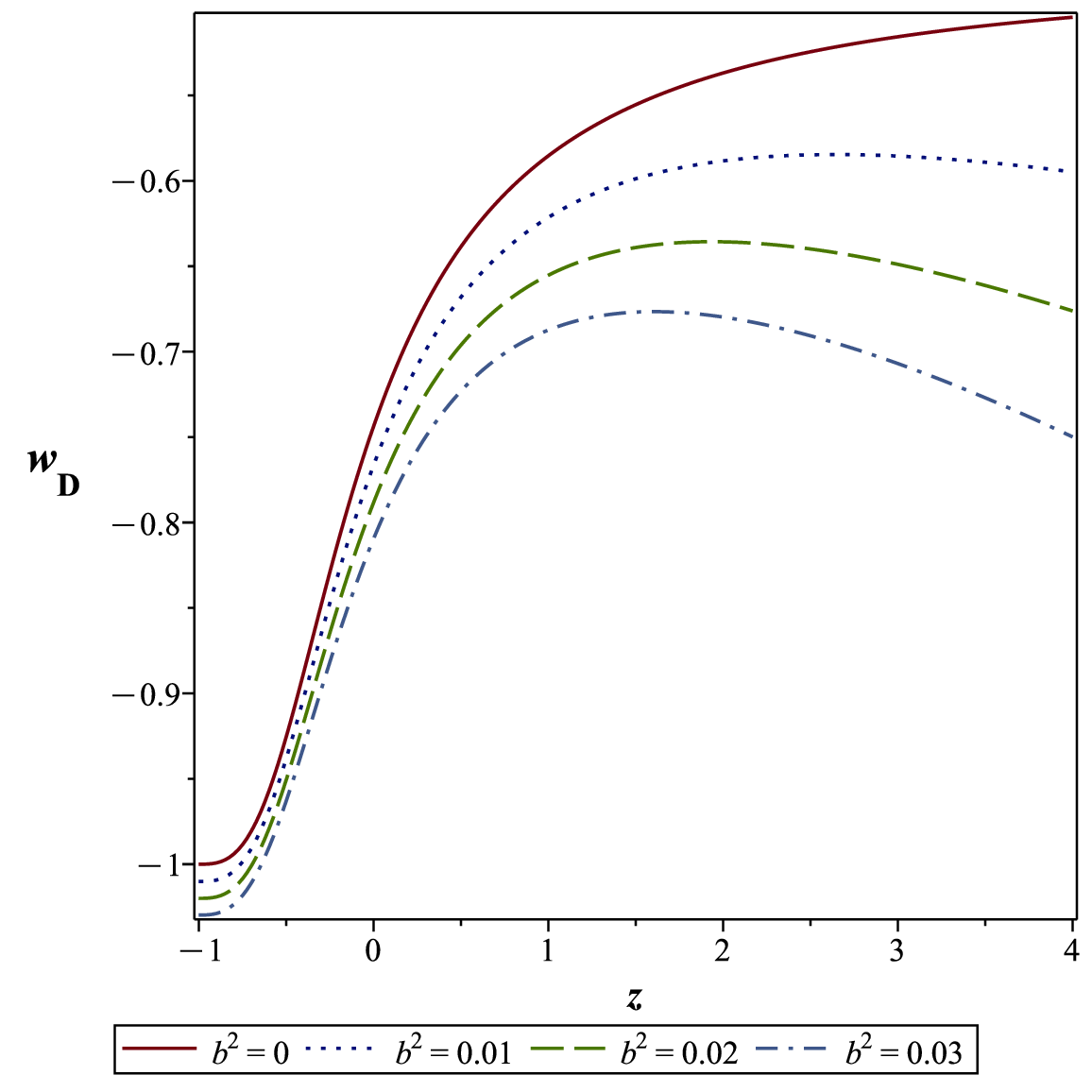}
\vspace{2mm}
\caption{Evolution of $w_D$ parameter for GDE in Barrow cosmology. In upper panel we set $b^2=0.02$ while in lower panel we take $\Delta=0.1$.} \label{wfbbvar}
\end{figure}

\begin{figure}
\centering
\includegraphics[scale=0.25]{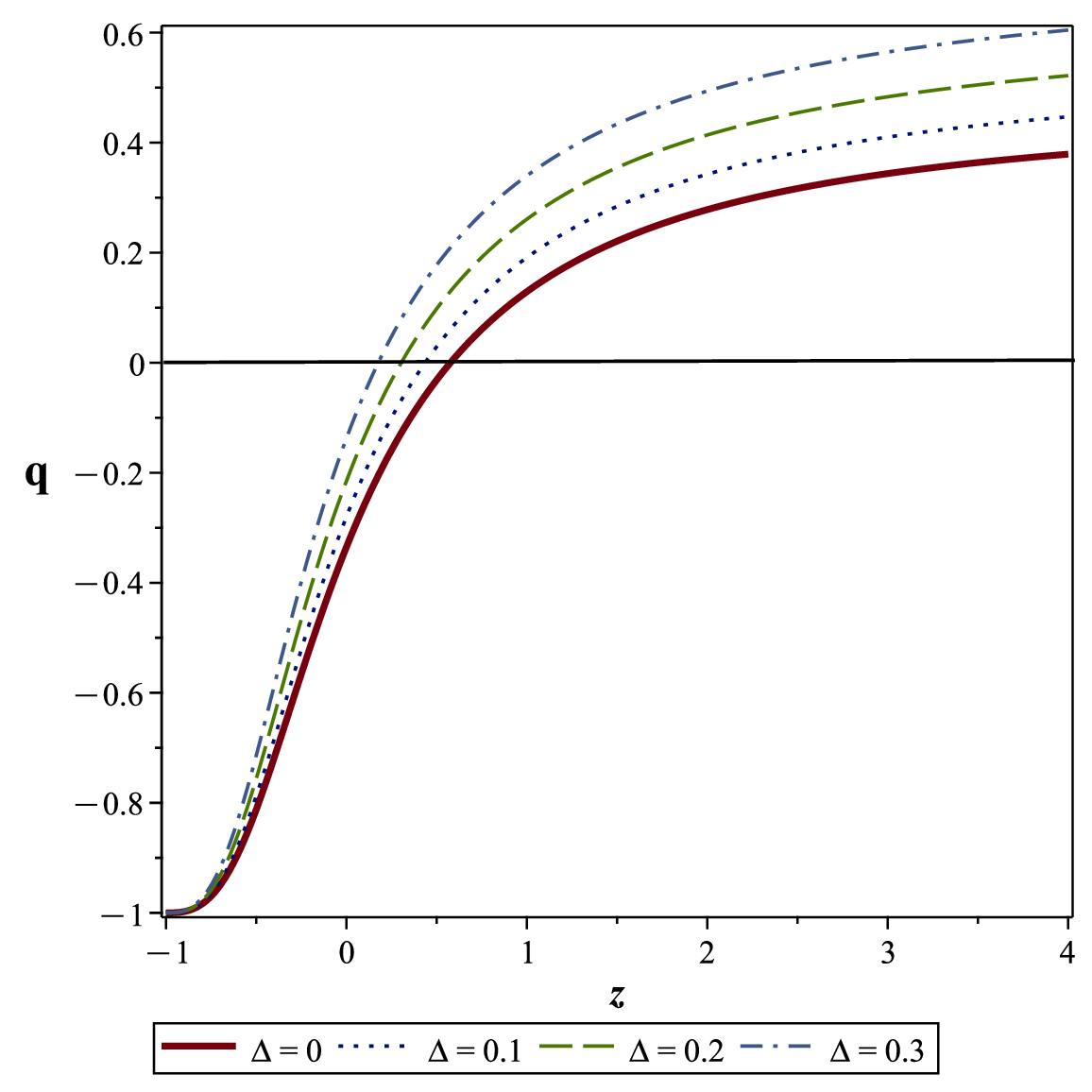}
\includegraphics[scale=0.25]{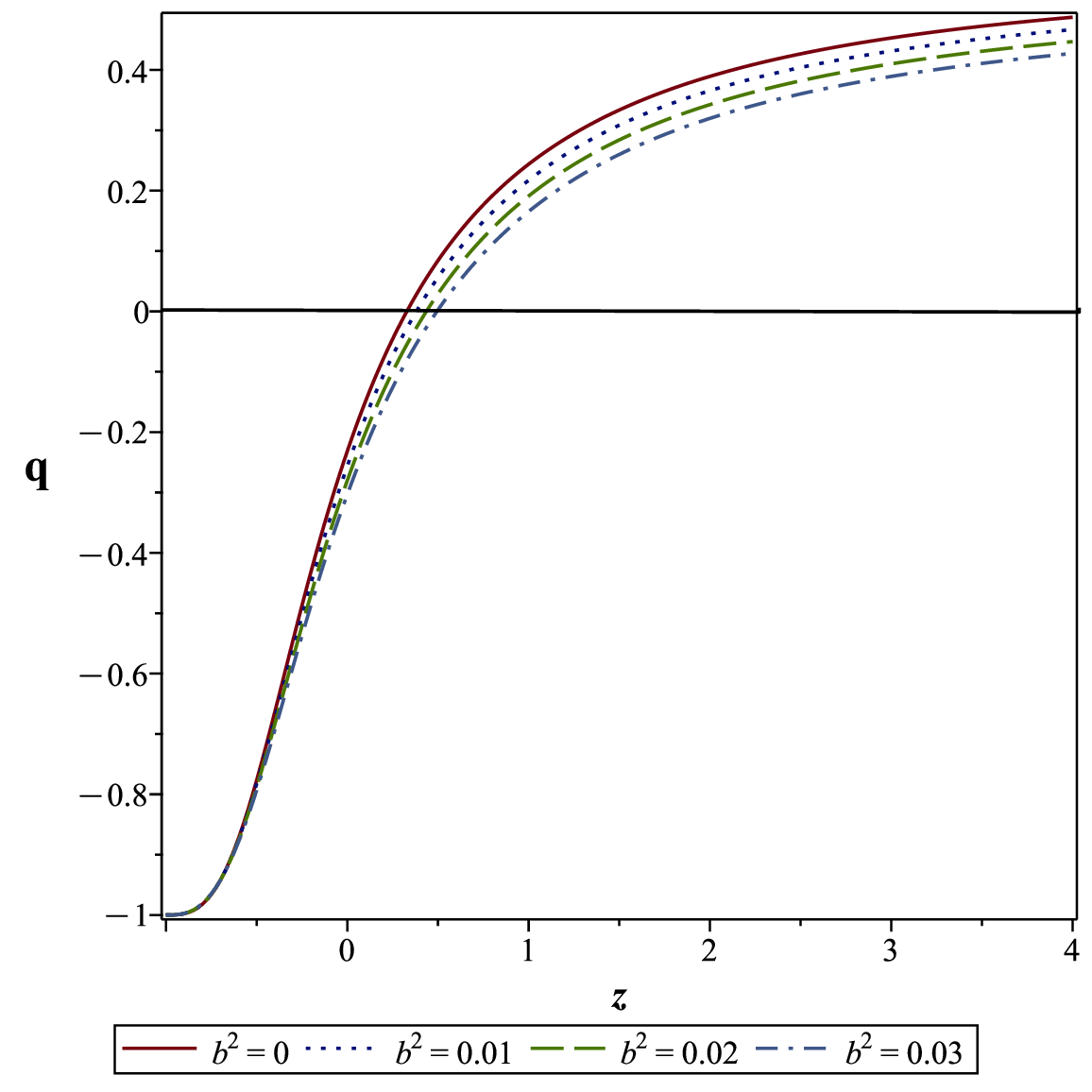}
\vspace{2mm}
\caption{Evolution of $q$ parameter for GDE in Barrow cosmology. In upper panel we set $b^2=0.02$ while in lower panel we take $\Delta=0.1$.} \label{qb}
\end{figure}

According to the new range of free parameters$(\Delta, b)$, obtained from previous task, we depict $\Omega_D$, $w_D$ and $q$ in Figs.
\ref{omegfbdelvar}-\ref{qb}. In these figures we set
$\Omega_{D0}=0.69$. According to Fig. \ref{omegfbdelvar} we
observe that for BGDE at early universe $\Omega_D\rightarrow0$
while the model predicts a phase of totally DE dominated
($\Omega_D\rightarrow 1$) in the future. This figure also reveals
that larger choices of $\Delta$ leaves a larger contribution of
GDE at early epochs. Lower panel of Fig. \ref{omegfbdelvar} shows
the impact of adding an interaction to evolution of $\Omega_D$.
Fig.\ref{wfbbvar} displays evolution of the $w_D$. This figure
implies that increasing $\Delta$ significantly affects the
behavior of the GDE, specially at higher redshift. Lower
panel of this figure states that introducing an interaction term
in the Barrow cosmology leads the GDE model to cross the phantom
line($w=-1$), favored by observations, at late epochs while this
behavior is absent in non-interacting GDE case in Barrow cosmology
signaling that the fate of universe will end with a big rip
singularity. This behavior is similar to what already presented
in the literature for GDE in standard
cosmology\cite{Sheykhi:2011xz,Cai:2010uf}. It could be seen that
$w_D$ will almost be independent of $\Delta$ at future approaches
a same limit $w_D<-1$ depending on $b^2$.

Figure \ref{qb}, shows that introducing non-extensive behavior to
the GDE leads a delay in the time of transition from deceleration
to acceleration($z_t$). According to this figure the GDE in barrow
cosmology with $\Delta=0.1$ and $b^2=0$ has $z_t=0.24$ while
taking an interaction term with $b^2=0.03$ leads $z_t=0.50$. Thus
one can conclude that an interacting Barrow GDE is more favored by
observations\cite{Planck:2018vyg}.

In facing any model of DE we expect a stable and clustered
universe at present epoch. Squared sound speed($v^{2}_s$), is a
key quantity which reveals signals of instability against
perturbations in a semi-Newtonian approach. In a short wavelength
limit the perturbations in the cosmic background could be as
\begin{equation}\label{delta}
    \delta\propto e^{\pm i\omega t},
\end{equation}
where $\omega\propto v_s$. The negative $v_s^2$, can lead to a
growing mode of perturbations and make the cosmic fluid unstable.
However, the positive sign of the $v_{s}^2$ has an oscillating
behavior which implies a stable behavior against perturbations. So
the sign of the squared sound speed is of special importance in
the stability issue. The definition of the adiabatic squared sound
speed reads
\begin{equation}\label{v2def}
    v_s^2=\frac{dP}{d\rho}.
\end{equation}
In order to obtain the squared sound speed in a background filled with matter and DE components the definition could be written as \cite{Cai:2009ph}
\begin{equation}\label{v2def2}
    v_s^2=\frac{dP}{d\rho}=\frac{\dot{P}}{\dot{\rho}}=\frac{\dot{w}_D\rho_D+w_D\dot{\rho}_D}{\dot{\rho}_D+\dot{\rho}_m}.
\end{equation}
After a little algebra and using Eqs.(\ref{omegaev}), (\ref{wd1}) we reach
\begin{equation}\label{barrowv}
v^{2}_s=\frac{(1-\Delta)\Omega^{2}_D+(b^2(3-2\Delta)+\Delta-1)\Omega_D-b^2(2-\Delta)^2}{(\Omega_D-2+\Delta)^2}.
\end{equation}

\begin{figure}
\centering
\includegraphics[scale=0.25]{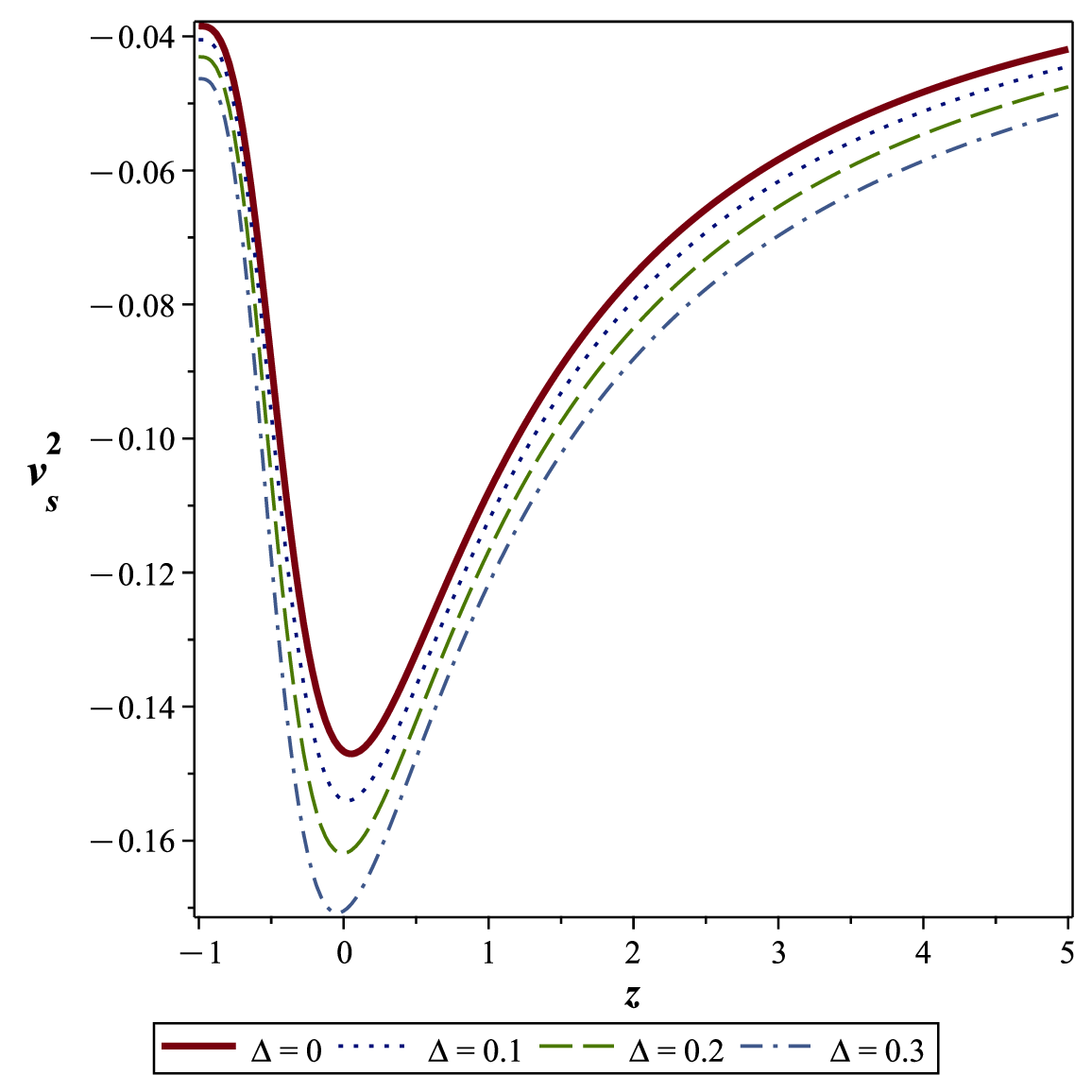}
\includegraphics[scale=0.25]{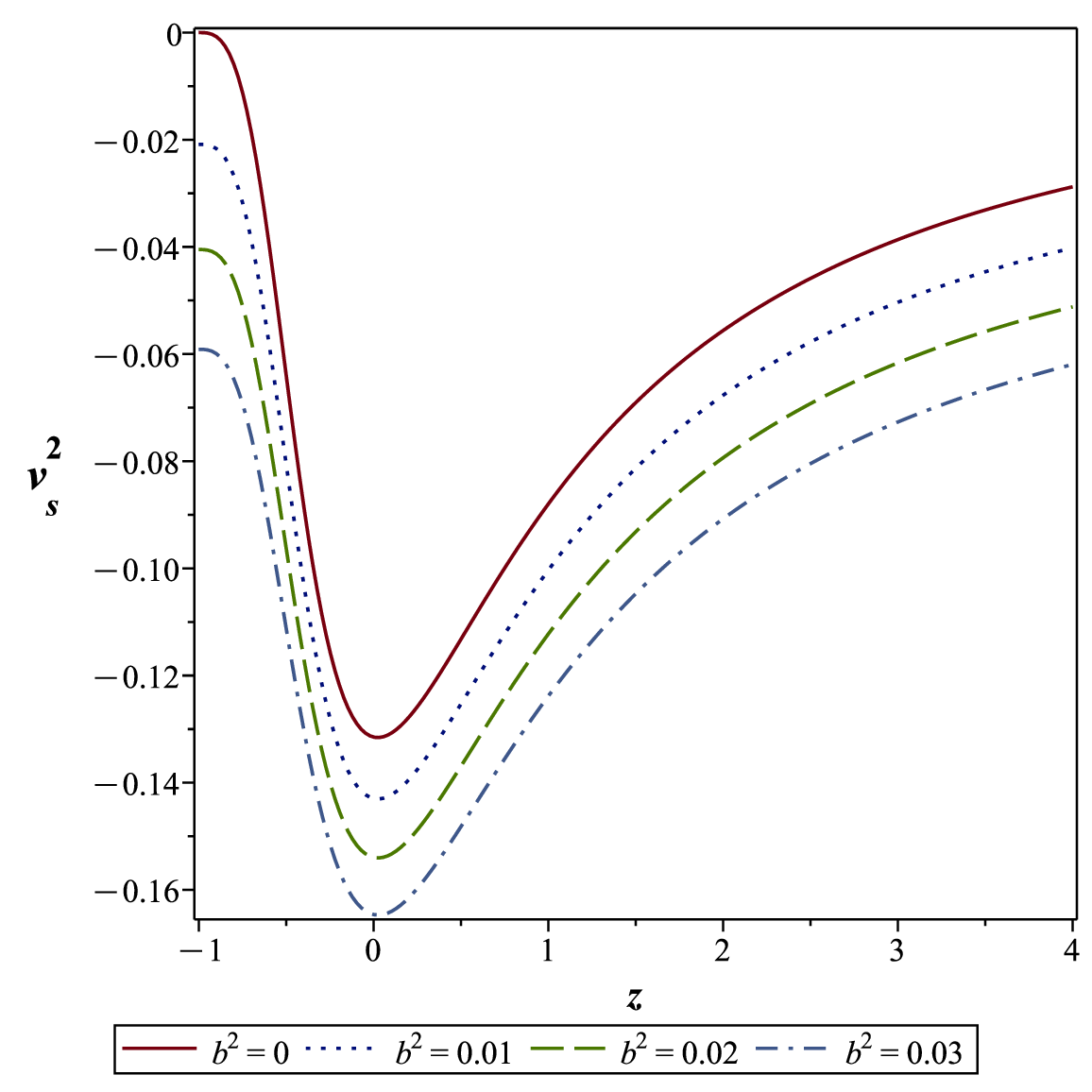}
\vspace{2mm} \caption{The chance of stability against
perturbations is discussed through plots of $v^{2}_s$ versus $z$.
In upper panel we set $b^2=0.02$ while in lower panel we take
$\Delta=0.1$.} \label{vs2b}
\end{figure}
It is worthy to note that above equation reduces to the
corresponding relation in the standard GR cosmology for
$\Delta=0$\cite{Cai:2010uf}. we plotted the squared sound speed
for different choices of $\Delta$ and $b^2$ in Fig.\ref{vs2b}. It
is clear that Barrow ghost dark energy(BGDE) is unstable against
perturbations for all choices of $\Delta$ and $b^2$. However it
can be seen from lower panel of Fig.\ref{vs2b} that for
$\Delta=0.1$ and $b^2=0$, the model meets the borderline at
future. In next step we consider the so-called statefinder
analysis in the case of BGDE to find the evolutionary status of
the model with respect to the $\Lambda$CDM. The statefinder
analysis introduced for first time in \cite{Sahni:2002fz}, where a
pair of dimensionless, geometrical
parameters,\{\textit{r},\textit{s}\} presented to discriminate
different DE models. These pair of parameters read
\begin{eqnarray}\label{rsdef}
r&=&\frac{\dddot{a}}{aH^3}=2q^2+q-q^{'}, \\
s&=&\frac{r-1}{3(q-\frac{1}{2})}.\nonumber
\end{eqnarray}label{rs}
where these parameters are background parameters since they are
constructed from scale factor ($a$) and its temporal derivatives.
For $\Lambda$CDM, in a flat background the statefinder pair take a
fixed value \{r, s\} = \{1, 0\}. Using above definitions we obtain
these parameters in the case of our study. The results are
\begin{eqnarray}\label{rbarrow}
r&=&-\frac{1}{(\Omega_D-2+\Delta)^3}\times\\
&\big[ &9\big(-\frac{\Delta^3}{9}+\big((1-\Omega_D)b^2-\Omega^{2}_D+\frac{1}{3}(5\Omega_D-1)\big)\Delta^2 \nonumber\\
&+&\big(\big(b^2+\frac{5}{3}\big)\Omega^{2}_D+\big(b^4-b^2-\frac{8}{3}\big)\Omega_D-2b^4+\frac{2}{3}\big)\Delta\nonumber\\
&-&\frac{10\Omega^{3}_D}{9}+\big(-4b^2+\frac{8}{3}\big)\Omega^{2}_D+\big(-3b^4+8b^2-\frac{7}{3}\big)\Omega_D\nonumber\\
&+&4b^2(b^2-1)+\frac{8}{9}\big) \big]\nonumber
\end{eqnarray}

\begin{eqnarray}\label{sbarrow}
&&s=\frac{-2\left(b^{2}+\Omega_D -1\right) }{\left(2 b^{2}-\Delta +\Omega_D \right) \left(\Omega_D -2+\Delta \right)^{2}}\times\\
&&\left(-\Omega^{2}_D+\left(\Delta  b^{2}-\Delta^{2}-3 b^{2}+2 \Delta +1\right) \Omega_D -2 \left(-2+\Delta \right) \left(b^{2}-\frac{\Delta}{2}\right)\right) \nonumber
\end{eqnarray}

\begin{figure}
\centering
\includegraphics[scale=0.25]{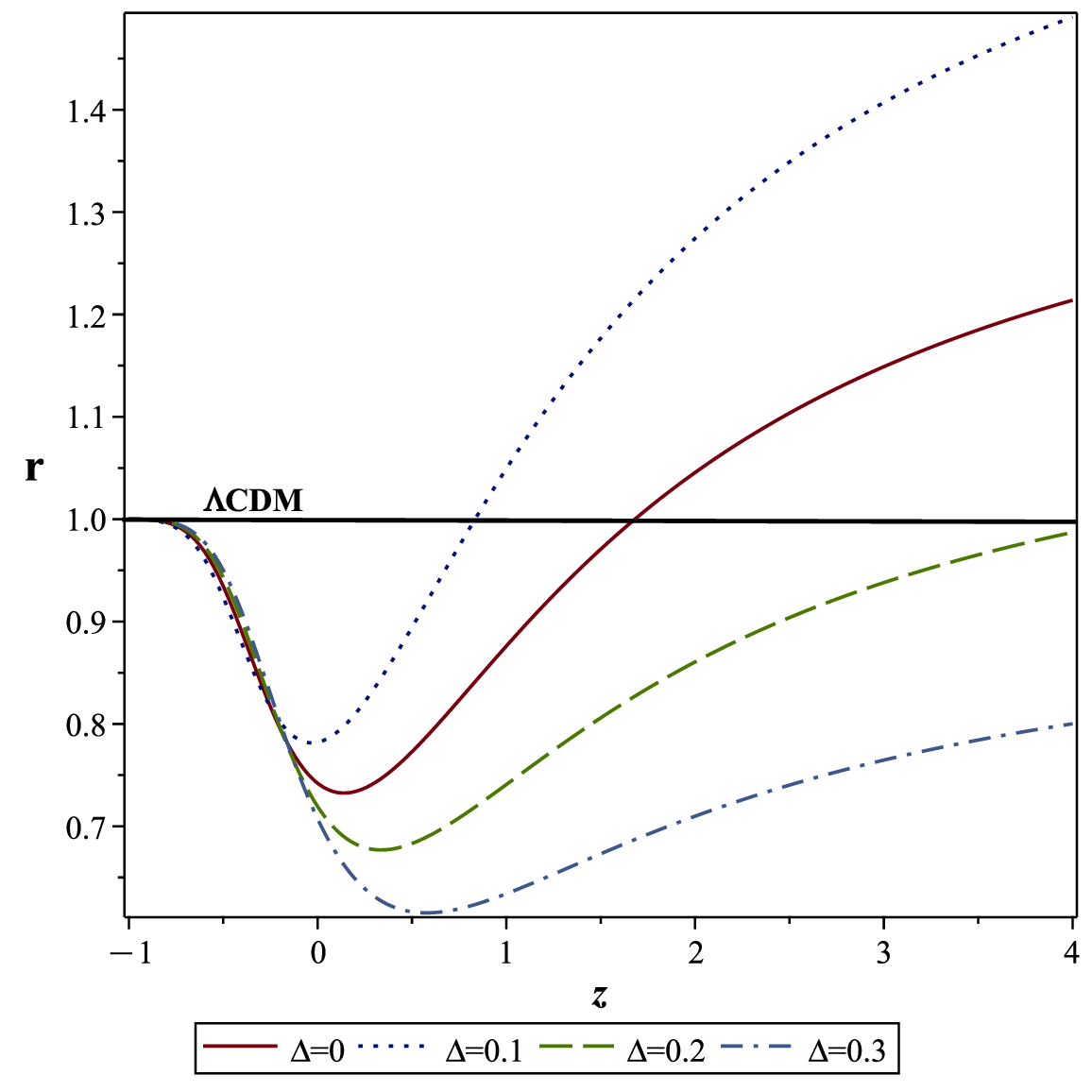}
\includegraphics[scale=0.25]{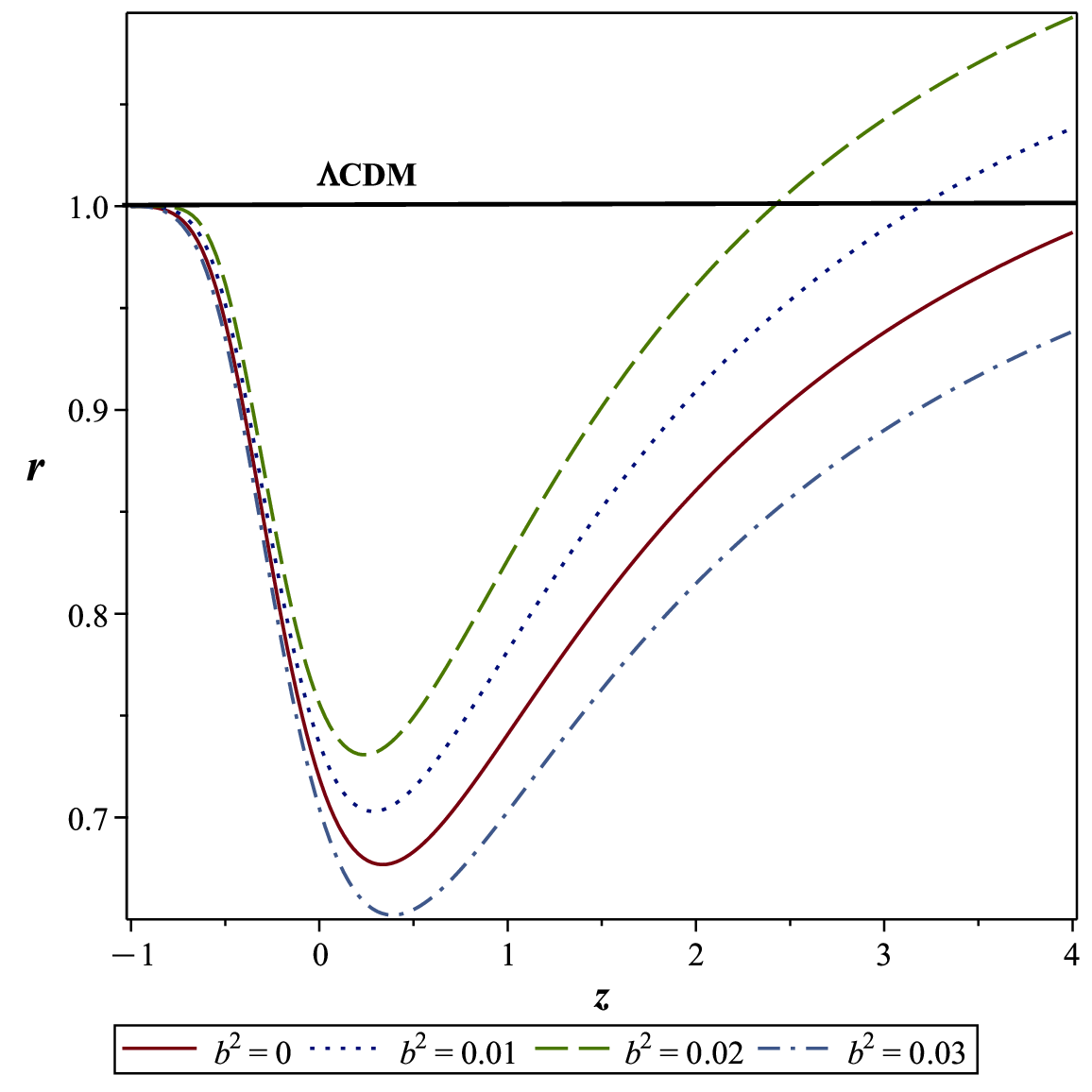}
\vspace{2mm}
\caption{Evolution of $r$ parameter for different choices of $\Delta$ and $b^2$.} \label{brvsz}
\end{figure}

\begin{figure}
\centering
\includegraphics[scale=0.25]{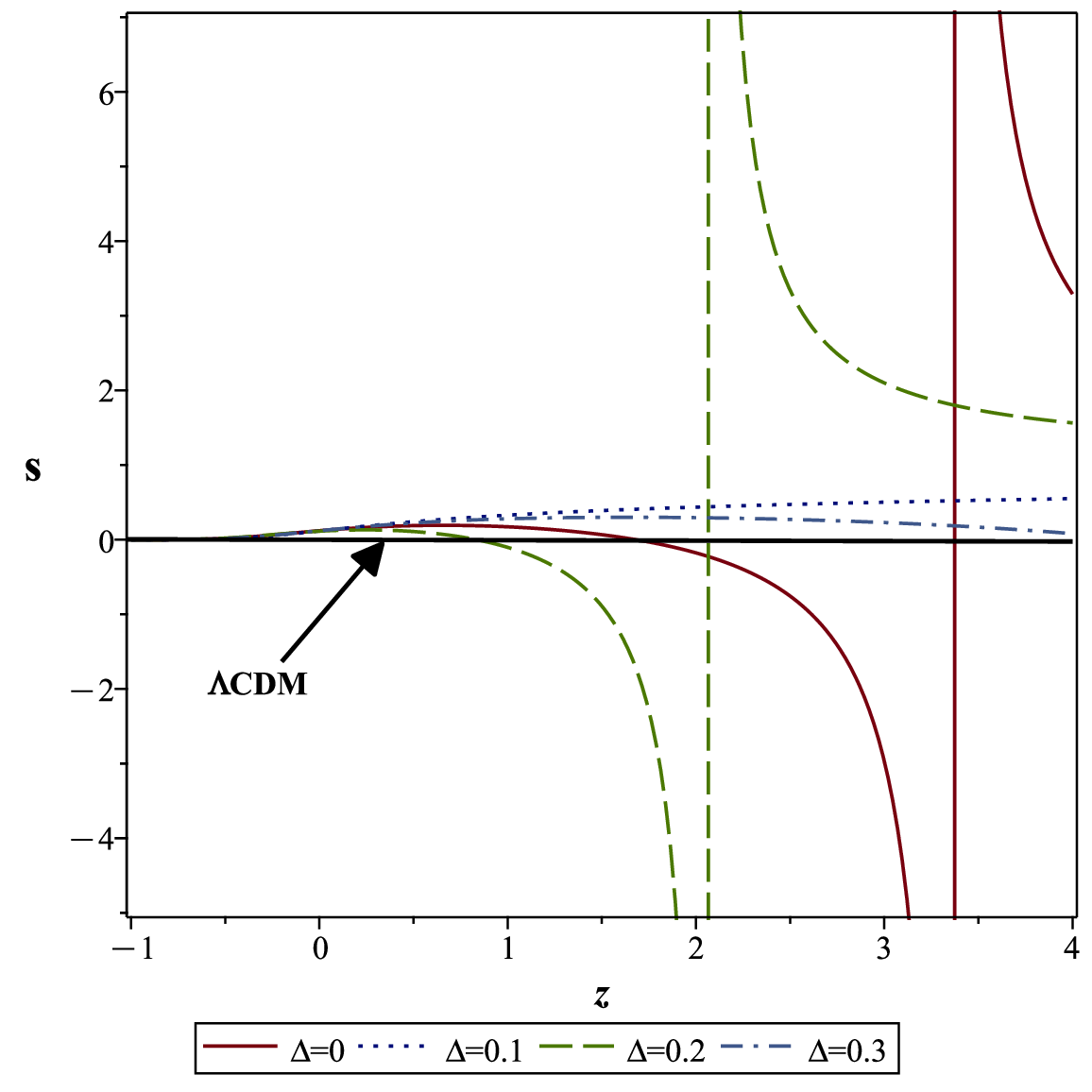}
\includegraphics[scale=0.25]{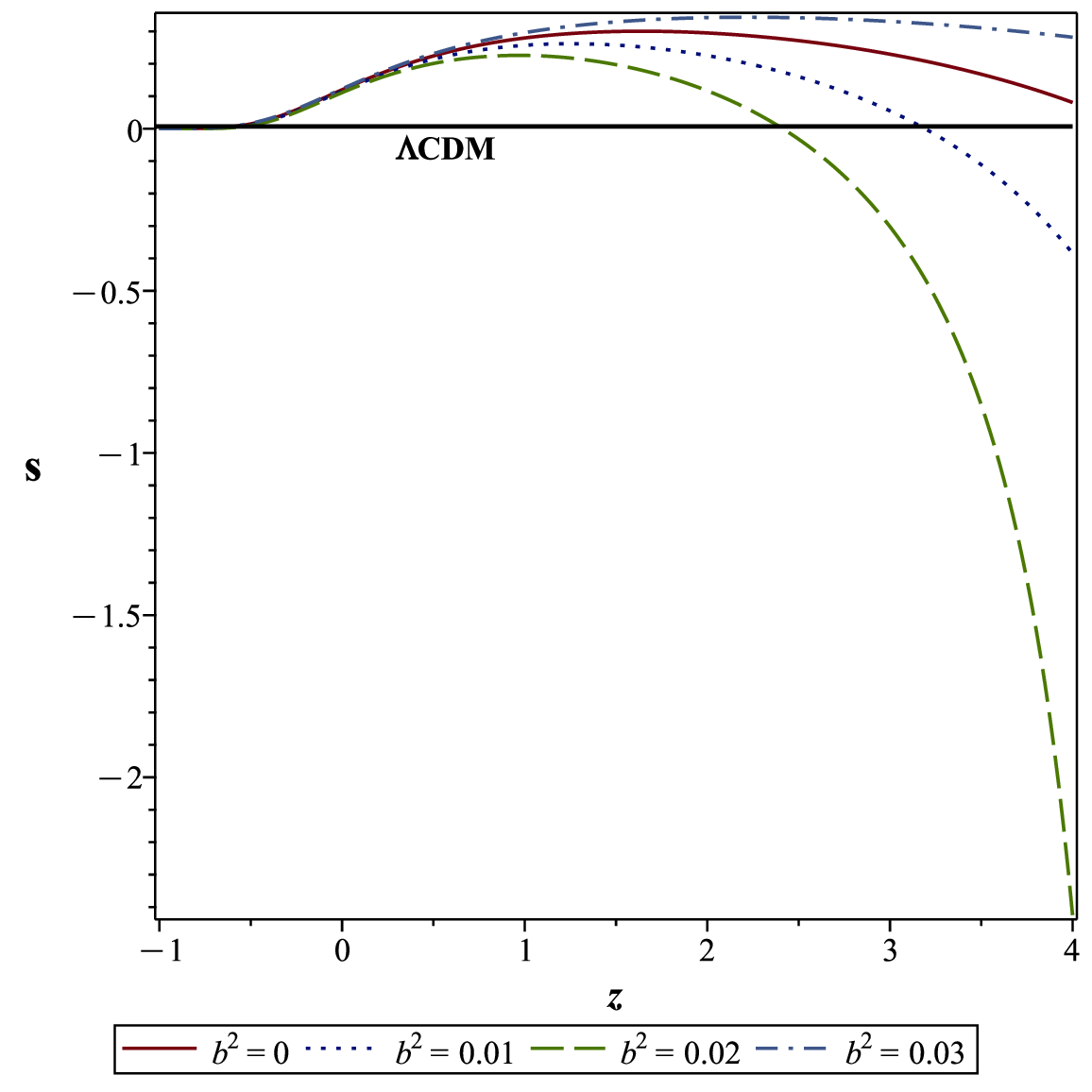}
\vspace{2mm}
\caption{Evolution of $s$ parameter for different choices of $\Delta$ and $b^2$.} \label{bsvsz}
\end{figure}

\begin{figure}
\centering
\includegraphics[scale=0.25]{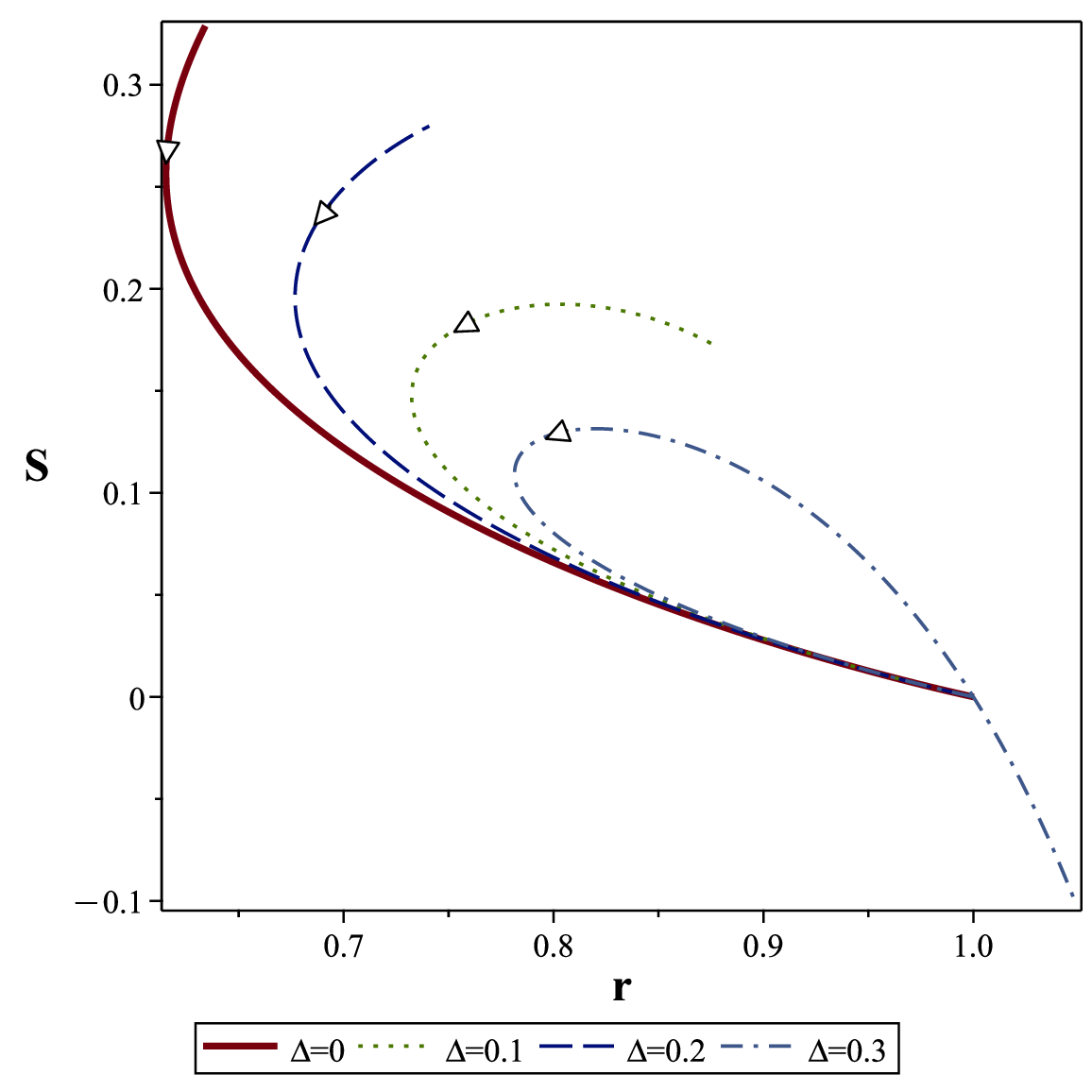}
\includegraphics[scale=0.25]{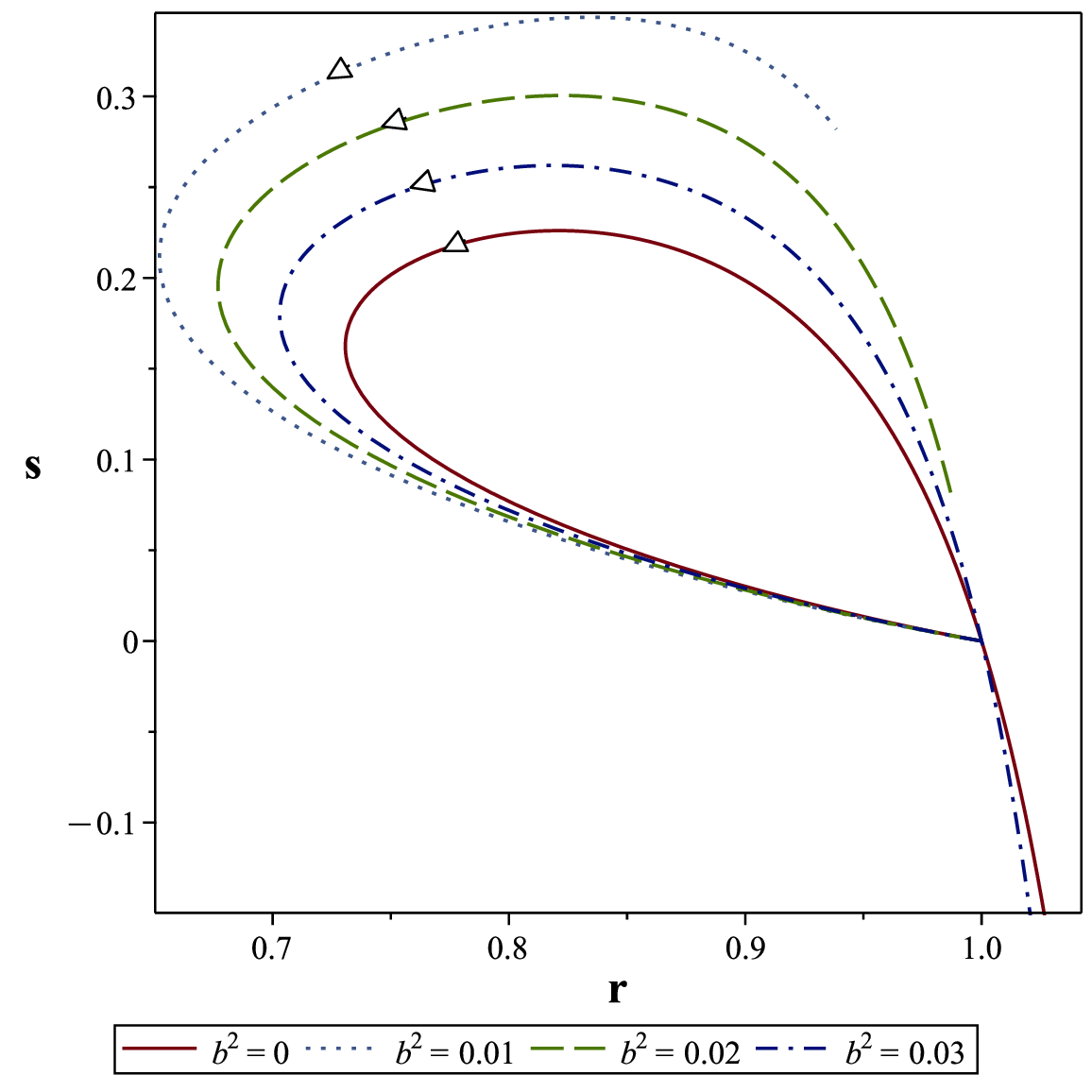}
\vspace{2mm}
\caption{The pattern of $\{s,r \}$ for different choices of $\Delta$ and $b^2$.} \label{bsvsr}
\end{figure}
Evolution of the statefinder parameters for the BGDE is plotted
against $z$ in Figs.\ref{brvsz}-\ref{bsvsr}. We find from these
figures that in all choices of $b^2$ and $\Delta$ the BGDE behaves
evidently distinct from the $\Lambda-CDM$. However in future BGDE
catches $\{r=1,s=0\}$ in all cases of interest. This analysis
reveals that BGDE starts from a matter dominated region while
enters the realm of standard cosmology in the future.

\section{Ghost dark energy in Tsallis cosmology} \label{Tsallis}
In this section we would like to consider cosmological
implications of TGDE. In this case the modified Friedmann equation
reads
\begin{equation}\label{tfriedmann}
\left(H^2+\frac{k}{a^2}\right)^{2-\beta}=\frac{8\pi
l^{2}_{P}}{3}\rho,
\end{equation}
where $\beta$ is the non-extensive Tsallis parameter, which is a
real parameter responsible of quantifying the degree of
non-extensivity. Necessity of $\gamma>0$ implies that $\beta<2$
\cite{Sheykhi:2018dpn}. Here, we work in a flat spacetime and thus
the Friedmann equation could be written as
\begin{equation}\label{tflatfriedmann}
H^{4-2\beta}=\frac{8\pi l^{2}_{P}}{3}\rho.
\end{equation}
Our universe is assumed to be filled with pressureless matter and
the GDE components ($\rho=\rho_m+\rho_D$). Once again we use a
same set of continuity equations (\ref{conseq}). Using Eqs.
({\ref{conseq}) and (\ref{tfriedmann}) and following same steps as
the previous section, one can obtain the EoS for the TGDE as
\begin{equation}\label{wsint}
w_D=\frac{\left(3-2 \beta \right) \Omega_D -2 b^{2} \left(-2+\beta
\right)}{\Omega_D  \left(\Omega_D -4+2 \beta \right)},
\end{equation}
where the interaction term in continuity equation is taken of the
form $Q=3b^2H(\rho_D+\rho_m)$. It is also a matter of calculation
to derive the evolutionary form of $\Omega_D$ as well as the
deceleration parameter $q$. The results are
\begin{equation}\label{omegaevs}
\Omega^{\prime}_D=-\frac{3 \Omega_D  \left(b^{2}+\Omega_D
-1\right) \left(-3+2 \beta \right)}{\Omega_D -4+2 \beta},
\end{equation}
\begin{equation}\label{qt}
q=\frac{3 b^{2}+2 \Omega_D -2 \beta +1}{\Omega_D -4+2 \beta}.
\end{equation}
It is worth mentioning that when $\beta\rightarrow1$, in both
interacting ($b\neq0$) and non-interacting ($b=0$) cases, the
above equations reduce to the corresponding equations in standard
cosmology \cite{Cai:2010uf}.

\subsection{Dynamical system analysis in Tsallis cosmology}
Here, as in the BGDE, we at first perform a same dynamical analysis in the case of Tsallis cosmology. The corresponding autonomous set of equations in this case read
\begin{equation}\label{omdprimes}
\Omega^{\prime}_D=\frac{g_{1dt}\Omega_{\mathrm{D}}^{2}+g_{2dt} \Omega_{\mathrm{D}}}{\Omega_{\mathrm{D}}-4+2 \beta},
\end{equation}

\begin{equation}\label{ommprimes}
\Omega^{\prime}_m=\frac{g_{1mt} \Omega_{m}^{2}+g_{2mt} \Omega_{m}+6 b^{2} \left(\frac{\Omega_{\mathrm{D}}}{2}-2+\beta \right) \Omega_{\mathrm{D}}}{\Omega_{\mathrm{D}}-4+2 \beta},
\end{equation}
where
\begin{eqnarray}
 g_{1dt}&=&-6 b^{2} \beta +9 b^{2}-8 \beta +13, \nonumber\\
 g_{2dt}&=&\left(-6 b^{2} \Omega_{m}-2 \Omega_{m}+8\right) \beta +9 b^{2} \Omega_{m}+4 \Omega_{m}-13,\nonumber\\
g_{1mt}&=& -6 \left(b^{2}+\frac{1}{3}\right) \left(\beta -2\right). \nonumber\\
 g_{2mt}&=&\left(\left(-6 \beta +15\right) \Omega_{\mathrm{D}}+6 \beta -12\right) b^{2}\nonumber\\ &+&\left(-8 \beta +13\right) \Omega_{\mathrm{D}}+2 \beta -4.\nonumber
 \end{eqnarray}
The admitted fixed points in this case are listed in Tab.(\ref{Tinttab}). The interesting point is that not in BGDE nor in TGDE the Barrow and Tsallis parameters ($\Delta$ and $\beta$) does not affect the number and existence of the fixed points. However, a closer look reveals the impacts of the models parameters on the stability and consequently on destination of the universe.

\begin{table}
\begin{tabular}{|c|c|c|c|c|c|}
  \hline
  fixed point & $\Omega_D$ & $\Omega_m$ & $q$ & Nature \\
  \hline\hline
  $P_1$ & 0 & 0 & $>0$ & saddle for $\beta>\frac{13}{8}$  \\
  &   &  & $ $ & unstable for $\beta<\frac{13}{8}$  \\ \hline
  $P_2$ & 0 & 1 & $>0$ & saddle for $\beta<\frac{3}{2}$\\
  &   &  & $ $ & stable for $\frac{3}{2}<\beta<2$  \\ \hline
  $P_3$ & $-b^2+1$ & $b^2$& $<0$ & stable for $\beta<\frac{3}{2}+\frac{1+b^2}{8}$ \\
  \hline
  \end{tabular}
  \caption{The admitted fixed points of TGDE in presence of an interaction term($Q=3b^2H(\rho_D+\rho_m)$). One should note that in this table we assumed $\beta<2$ according to \cite{Sheykhi:2018dpn}}
  \label{Tinttab}
\end{table}

\begin{figure}
\centering
\includegraphics[scale=0.35]{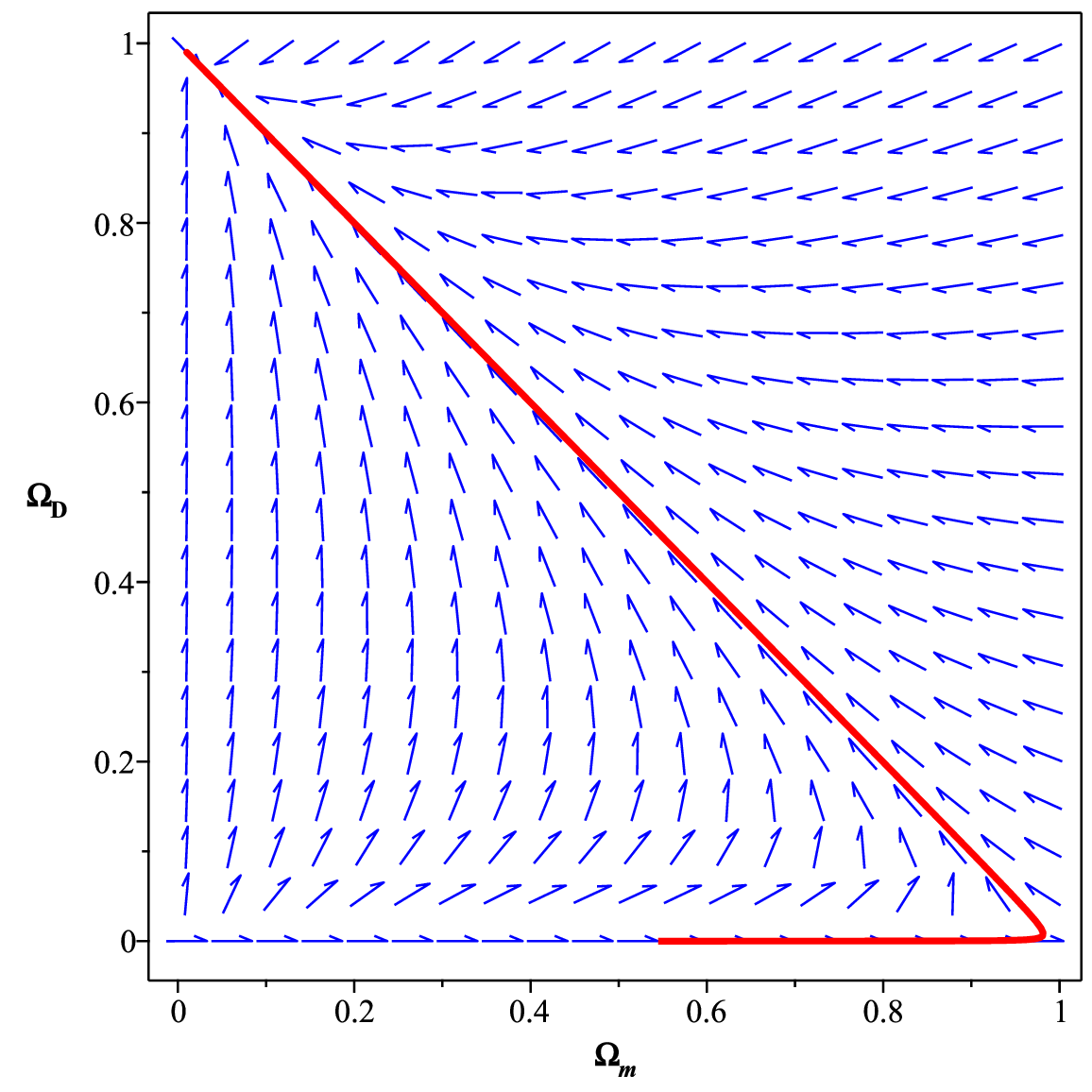}
\vspace{2mm}
\caption{Phase space evolution of the TGDE model. We set $b^2=0.01$ and take $\beta=1.1$. The solid red line identifies the universe with $\Omega_{D0}=0.69$ and $\Omega_{m0}=0.30996$} \label{tpspace}
\end{figure}
Once again the evolution pattern of the system starts from $P_{1}:
\Omega_{\mathrm{D}}=0,\Omega_{m}=0$ which corresponds to a
radiation dominated unstable phase of decelerated expansion.
$\frac{13}{8}<\beta<2$ leads to a saddle point while for
$\beta<\frac{13}{8}$, $P_1$ is an unstable point. In both cases
the system at this point shows a transient character. At this point the effective EoS parameter and deceleration parameter are $w_{eff}=-b^2$ and $q=\frac{3 b^{2}-2 \beta +1}{-4+2 \beta}$.

The second fixed point($P_2: \Omega_D=0, \Omega_m=1$) provides a
matter dominated epoch. The stability analysis
($\lambda_1=-3b^2-1,
\lambda_2=-\frac{3}{2}\frac{2\beta-3}{\beta-2}(b^2-1)$) reveals
that for $\frac{3}{2}<\beta<2$ and $b^2<1$, this point has an
stable nature and the universe will not transits to a future
attractor. However if $\beta<3/2$, then this point will be a
saddle point which has an unstable character and leaves the chance
of transition to future attractor. Thus necessity of transition
from a DM to DE dominated epoch puts a new constraint on $\beta$.
However one should note that this constraint is obtained in the
case $Q=3b^2 H(\rho_D+\rho_m)$. $P_3:
\Omega_D=-b^2+1,\Omega_m=b^2$ is another fixed point which
corresponds to a DE dominated epoch. Due to briefness we do not
mention the exact form of eigenvalues in this case. The
eigenvalues show that this point could be stable for
$\beta<(\frac{3}{2}+\frac{1+b^2}{8})$ which this condition is
already satisfied with the unstability condition of $P_2$. Thus if
the model evolves in to a DE dominated epoch it would be stable at
this point. At this stable point one can obtain $w_{eff}=-1=q$
which predicts a future big rip singularity as destination of the
universe. To see the validity of above description we plotted
phase space diagram in the TGDE case in Fig.(\ref{tpspace}).

As result we found that for GDE in Barrow and Tsallis cosmology
the system could result in a correct timeline order of evolution
with suitable choice of free parameters $\delta$ and $\beta$. The
resulting condition($\beta<3/2$) from dynamical system analysis
belongs to the allowed region($\beta<2$), mentioned in the
literature.

\subsection{Numerical description of TGDE}
\begin{figure}
\centering
\includegraphics[scale=0.25]{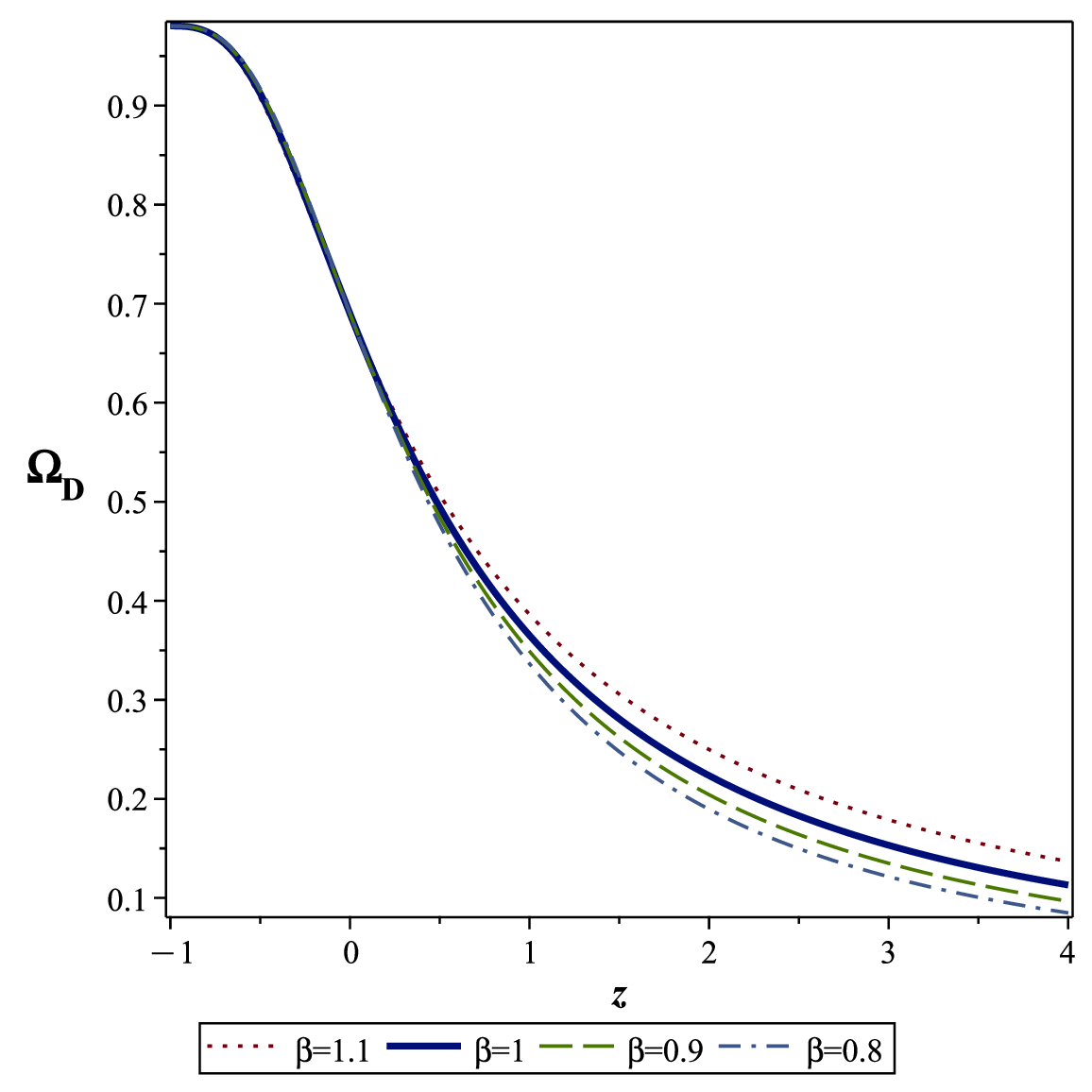}
\includegraphics[scale=0.25]{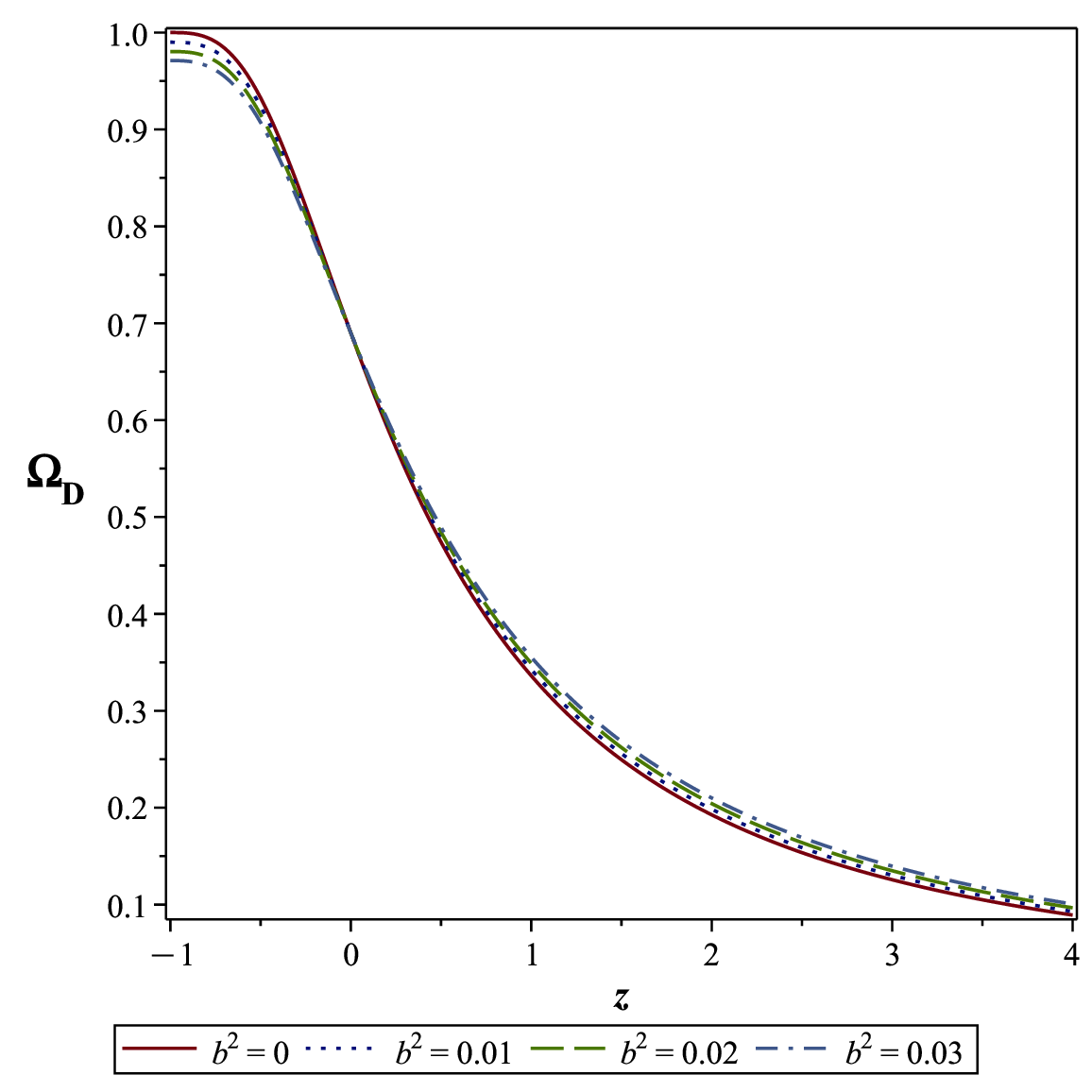}
\vspace{2mm}
\caption{Evolution of $\Omega_D$ parameter for GDE in Tsallis cosmology. In upper panel we set $b^2=0.02$ while in lower panel we take $\beta=0.9$.} \label{tomegf}
\end{figure}
In this section we also discuss a numerical description of the
TGDE. To this end, we plot the figures in allowed region of free
parameters obtained from the dynamical system analysis. According
to Fig. \ref{tomegf}, TGDE contribution to the cosmic content at
present and future is not very sensitive to $\beta$. However
larger values of $\beta$, result in a more contribution at early
epochs of universe. This result maybe could taken as an early DE
component which is interesting in curing the cosmic tensions. On
the other side lower part of Fig.\ref{tomegf} shows that
increasing $b^2$ results in larger values of $\Omega_D$ at far
future.

\begin{figure}
\centering
\includegraphics[scale=0.25]{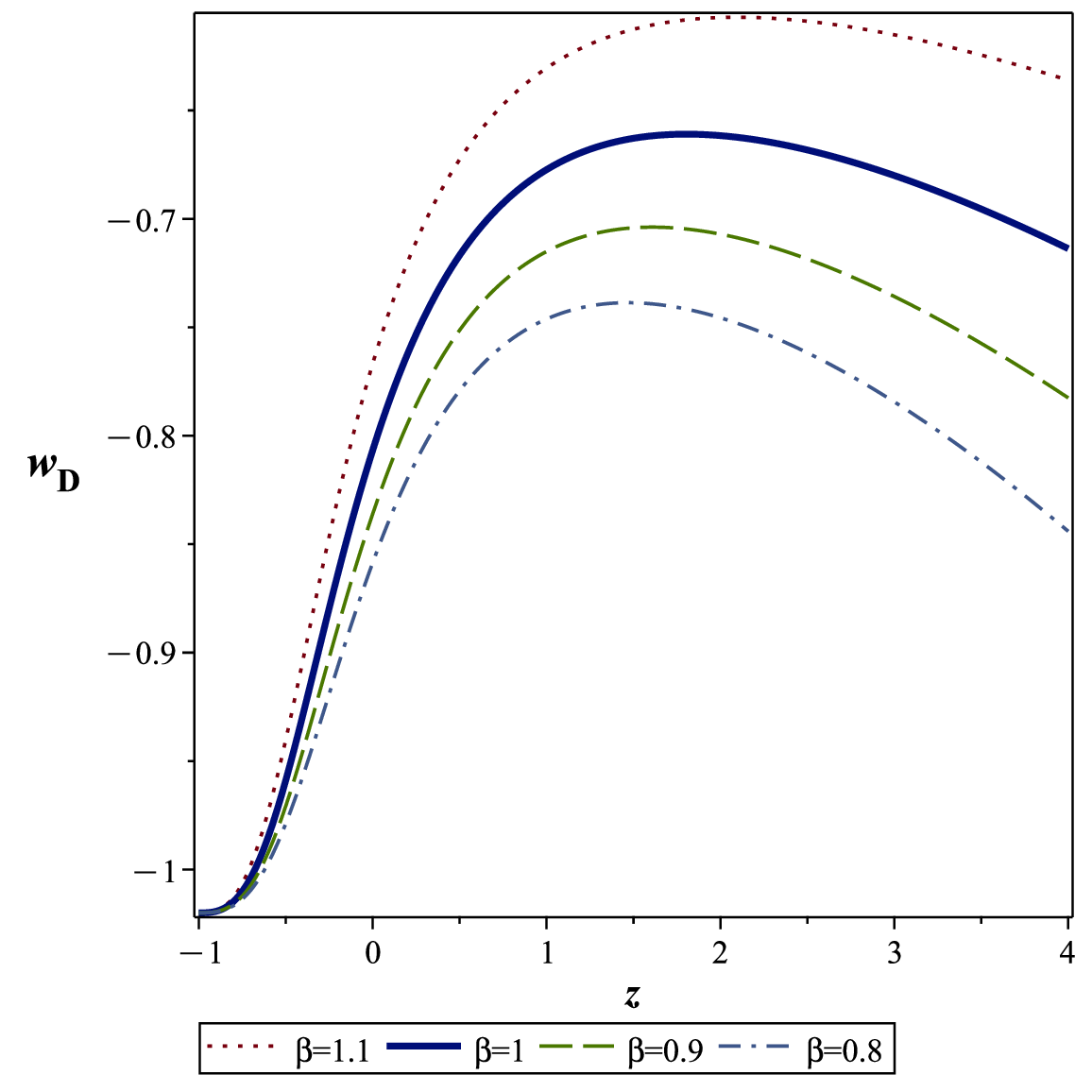}
\includegraphics[scale=0.25]{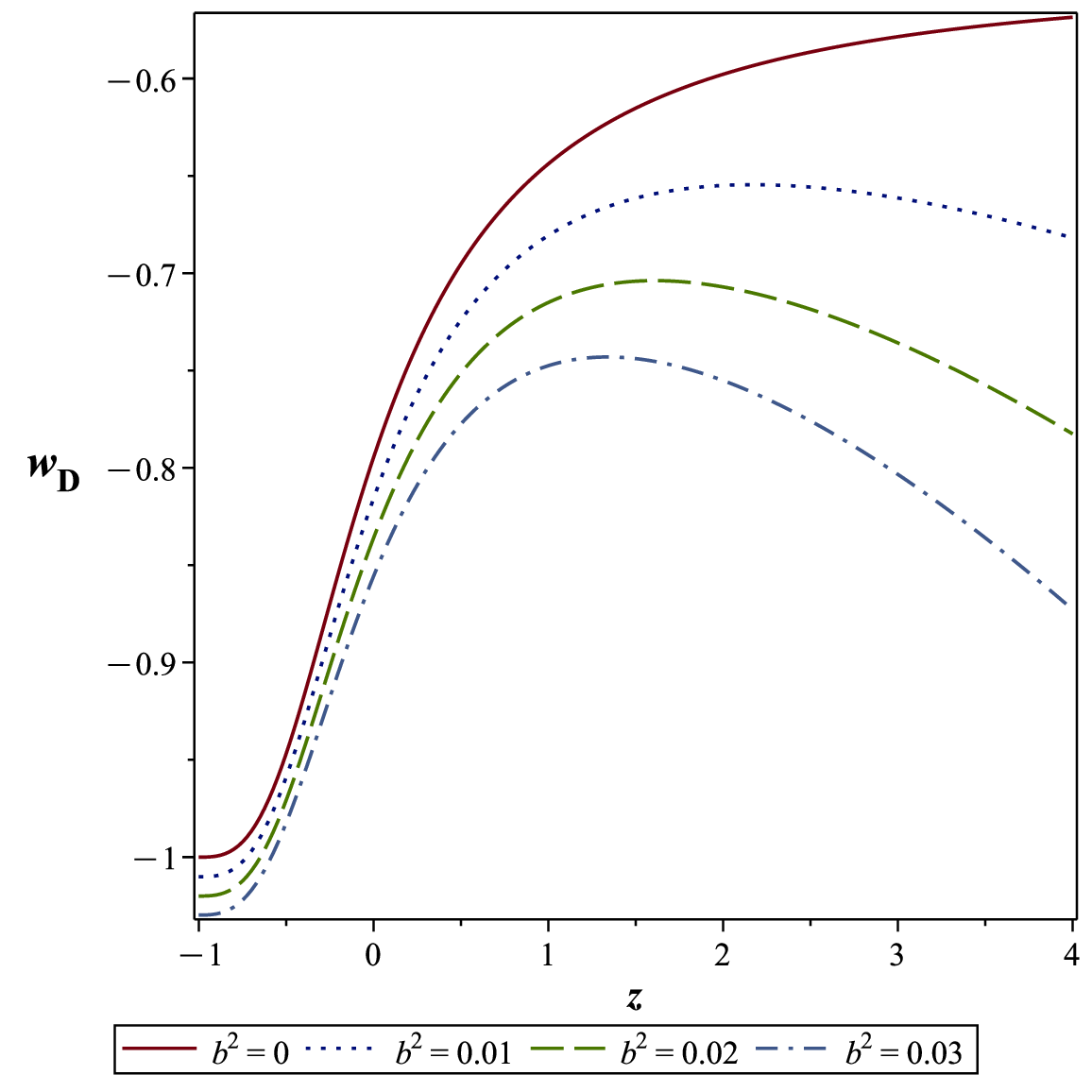}
\vspace{2mm}
\caption{Evolution of $w_D$ parameter for GDE in Tsallis cosmology. In upper panel we set $b^2=0.02$ while in lower panel we take $\beta=0.9$.} \label{twf}
\end{figure}
Fig. \ref{twf} indicates that with increasing the non-extensive
nature (increasing $\beta$) of the model decreases the value of
$w_D$. Besides introducing interaction between dark components,
$w_D$ will cross the phantom line, indicating a future big
rip singularity fate of universe, which is similar to the GDE
behavior in standard Friedmann cosmology\cite{Cai:2009ph}.
\begin{figure}
\centering
\includegraphics[scale=0.25]{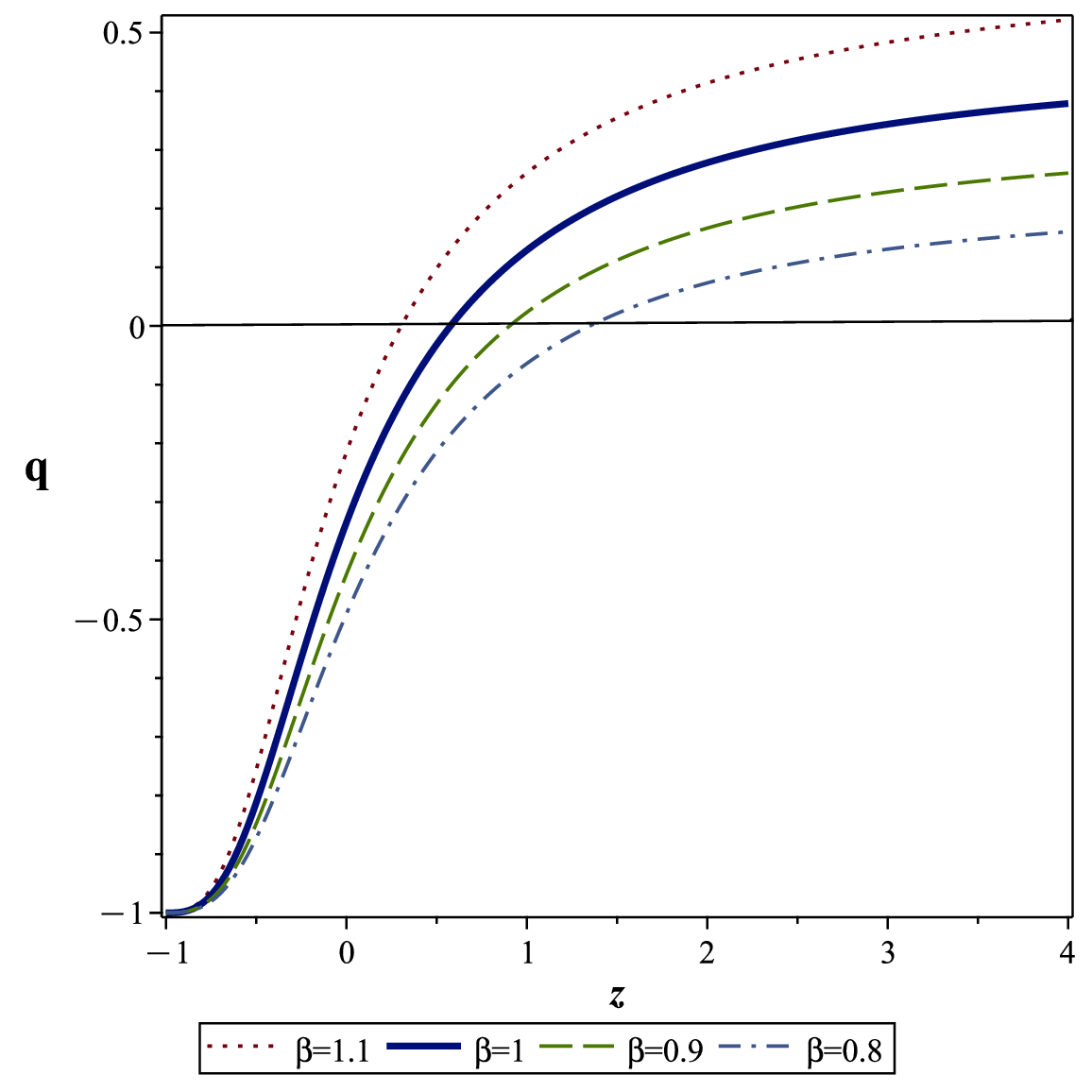}
\includegraphics[scale=0.25]{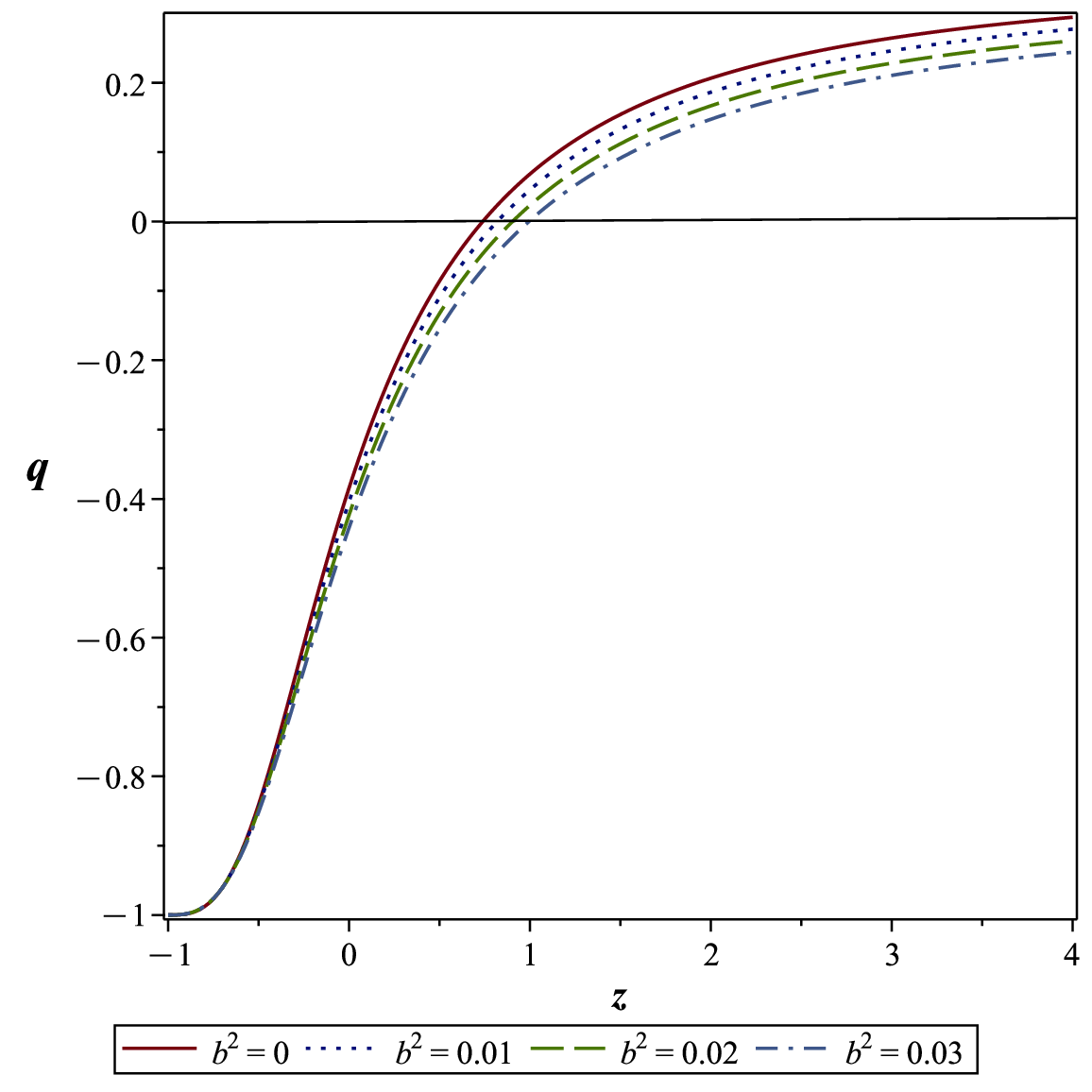}
\vspace{2mm}
\caption{Evolution of $q$ parameter for GDE in Tsallis cosmology. In upper panel we set $b^2=0.02$ while in lower panel we take $\beta=0.9$.} \label{tqf}
\end{figure}
The behavior of $q$ is depicted against $z$ in Fig.\ref{tqf}.
According to the upper panel of this figure, transition to an
accelerated expansion could happen during $0\leq z\leq1$. Since
the transition to a acceleration phase according to observation
should happen around $z\sim0.6$ \cite{Daly:2007dn,WMAP:2010qai}
thus the non-extensive parameter maybe should be close to unity.
From lower panel one finds that increasing $b^2$ leads to an
earlier phase transition from deceleration to acceleration. In
this figure the phase transition occurs at the observational
favored interval $0.7\leq z \leq 1$ for $\beta=0.9$ and
$b^2\in[0,0.03]$. One can also find in both panels of the
Fig.\ref{tqf} that the deceleration parameter approaches $q=-1$
which resembles a phase of de-Sitter expansion.

We also would like to explore the stability of the GDE in Tsallis
cosmology. The squared sound speed $(v^2_{s})$ in this case in
presence of an interaction between dark components read
\begin{eqnarray}
&&v^{2}_{s}=\frac{1}{\left(\Omega_D -4+2 \beta \right)^{2}}\times\\
&&[\left(3-2 \beta \right) \Omega^{2}_{D}+\left(-4 b^{2} \beta +7 b^{2}+2 \beta -3\right) \Omega_{D} \nonumber\\&&-4 b^{2} \left(-2+\beta \right)^{2}].\nonumber
\end{eqnarray}

\begin{figure}
\centering
\includegraphics[scale=0.25]{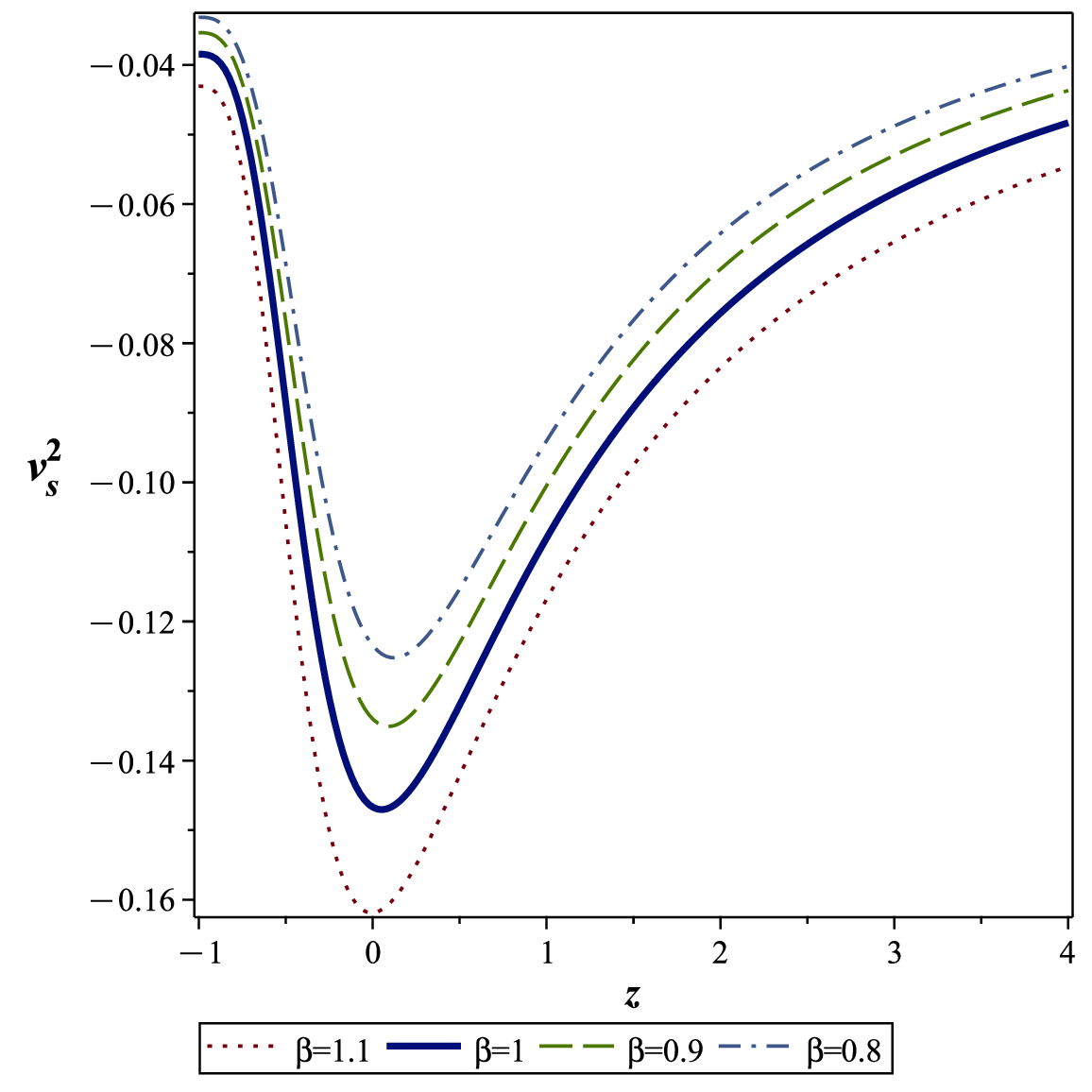}
\includegraphics[scale=0.25]{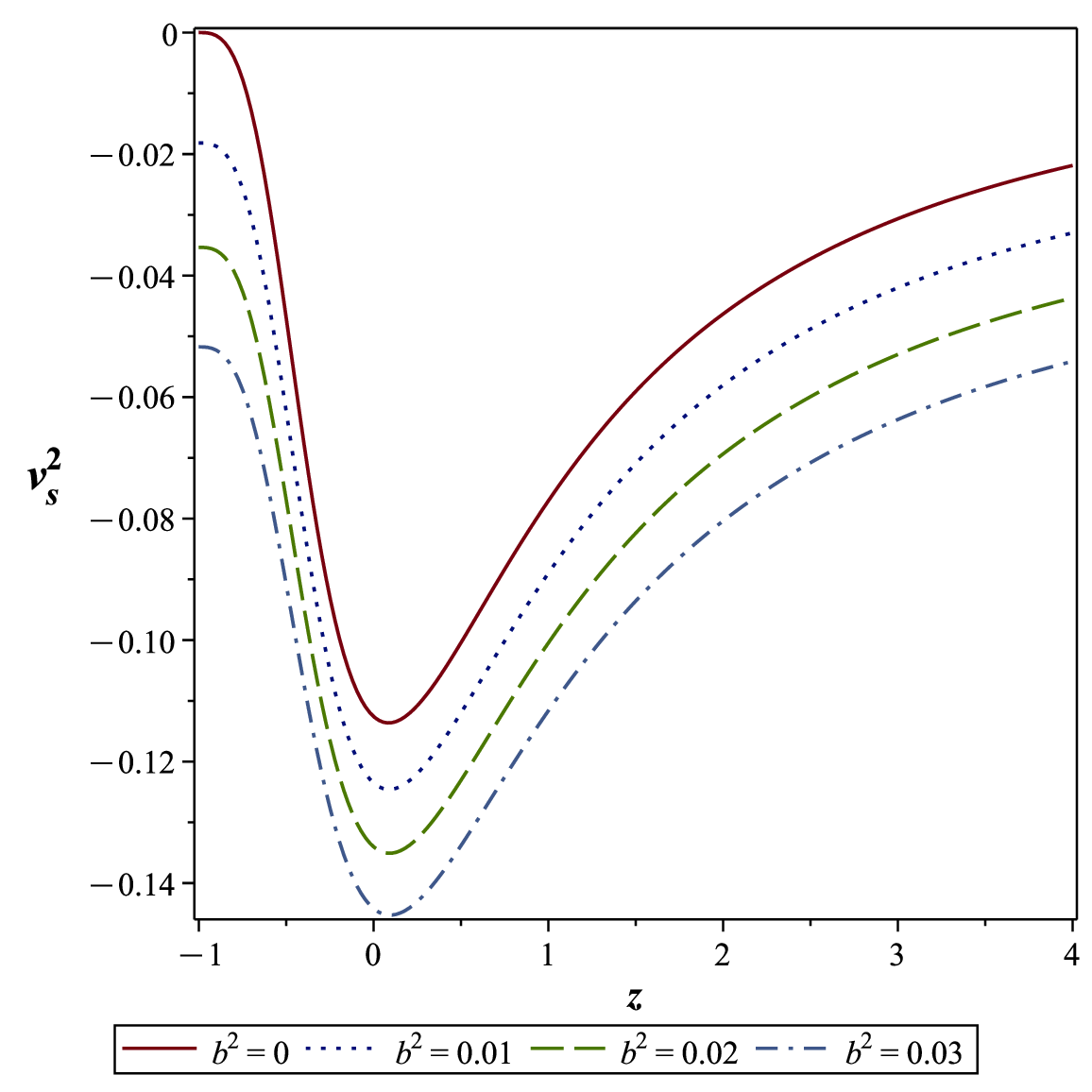}
\vspace{2mm}
\caption{Evolution of $v^{2}_{s}$ parameter for GDE in Tsallis cosmology. We set $b^2=0.02$ in upper panel and take $\beta=0.9$ in the other panel.} \label{tvs2f}
\end{figure}
In Figs.\ref{tvs2f} we depicted $v^{2}_{s}$ for different choices
of $\beta$ and the coupling constant $b^2$. This figure reveals
that increasing $\beta$(or $b^2$) leads more negative values for
$v^{2}_{s}$. Further this figure shows that altering the
non-extensive parameter $\beta$ can not change the unstable nature
of the model and the TGDE is still suffering the stability issue.
\begin{figure}
\centering
\includegraphics[scale=0.25]{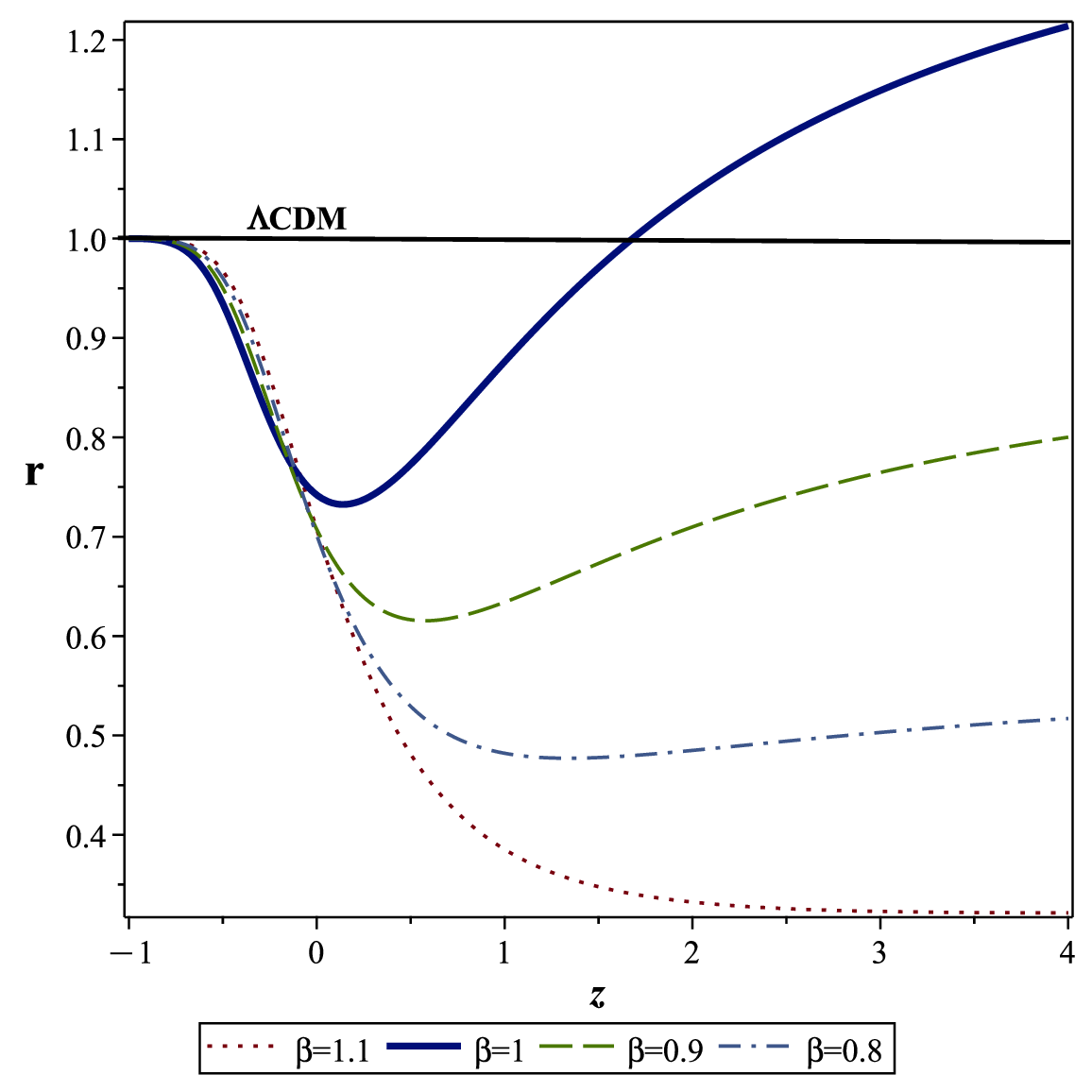}
\includegraphics[scale=0.25]{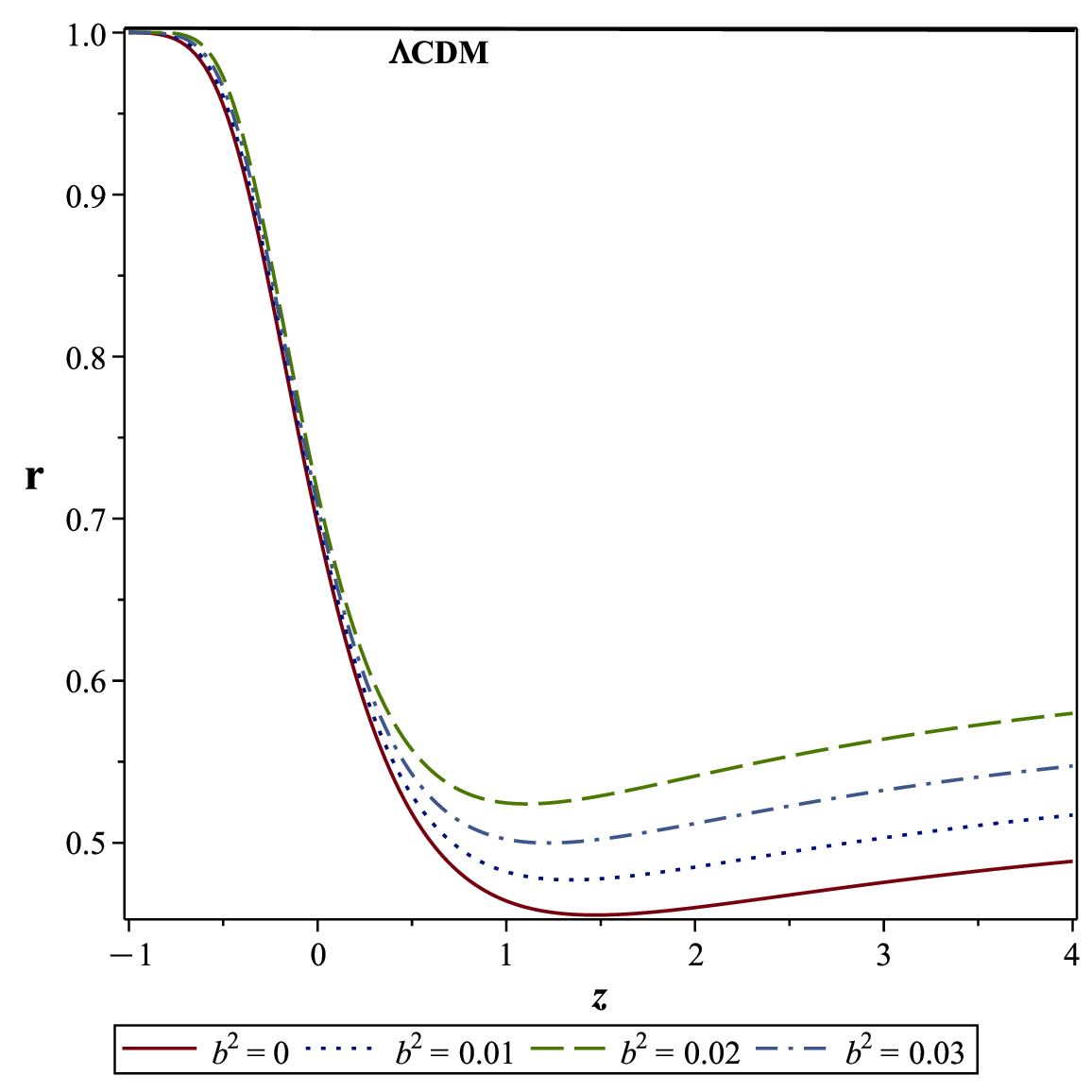}

\vspace{2mm}
\caption{In this figure we depicted the evolution of $r$ parameter against redshift. In upper panel we set $b^2=0.02$ and in the lower part we take $\beta=0.9$.} \label{tstrf}
\end{figure}

\begin{figure}
\centering
\includegraphics[scale=0.25]{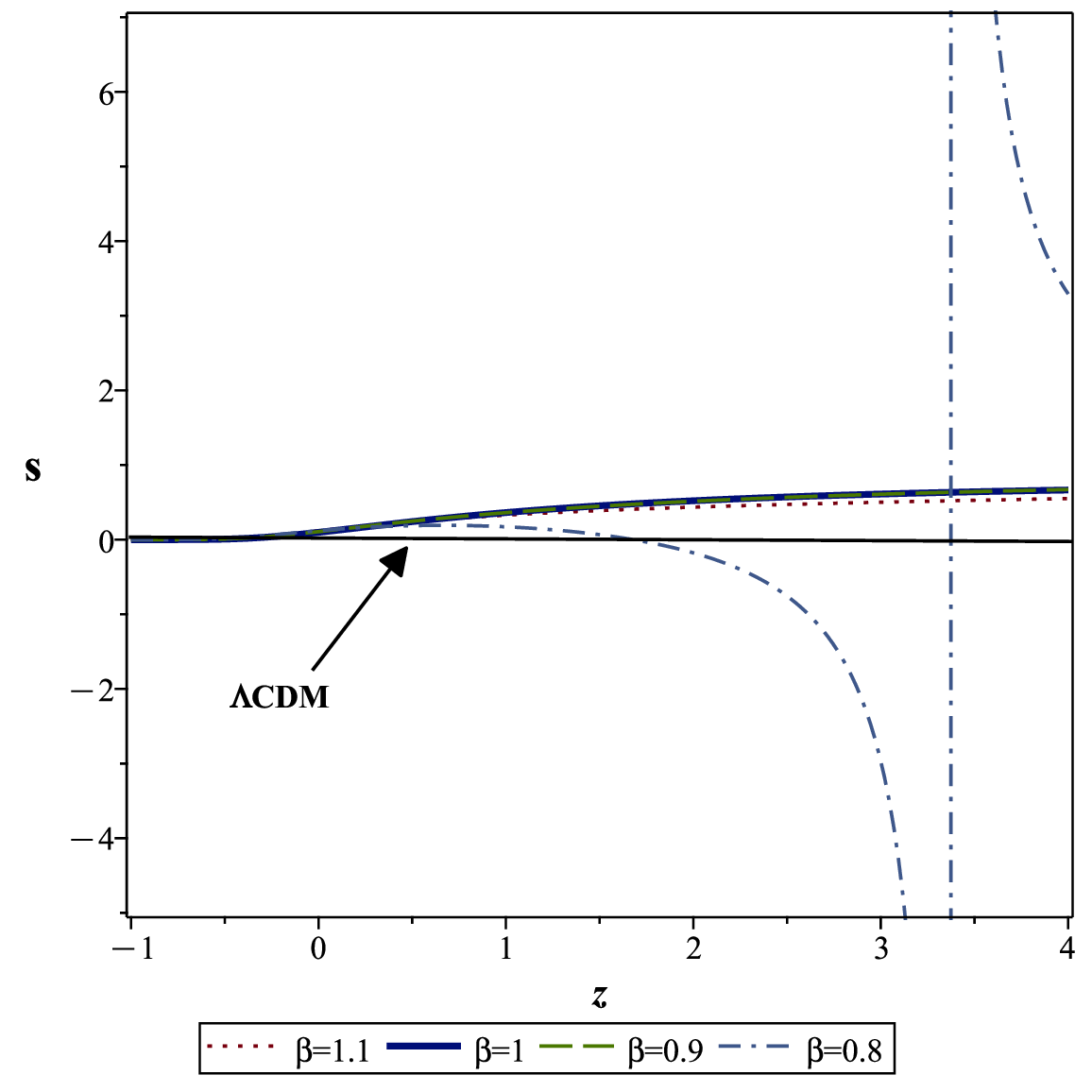}
\includegraphics[scale=0.25]{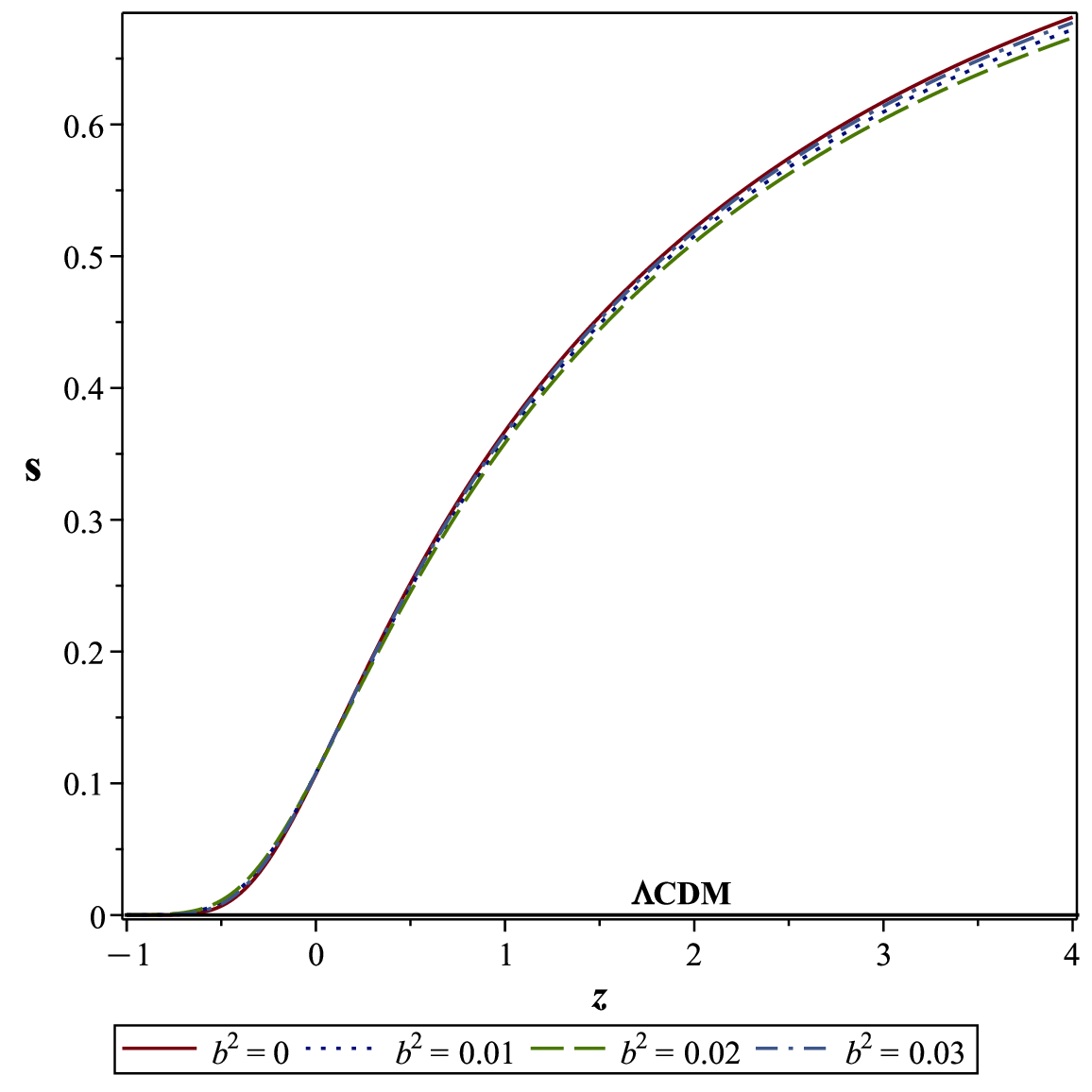}

\vspace{2mm}
\caption{In this figure we depicted the evolution of $s$ parameter against redshift. In upper panel we set $b^2=0.02$ and in the lower part we take $\beta=0.9$.} \label{tstsf}
\end{figure}

\begin{figure}
\centering
\includegraphics[scale=0.25]{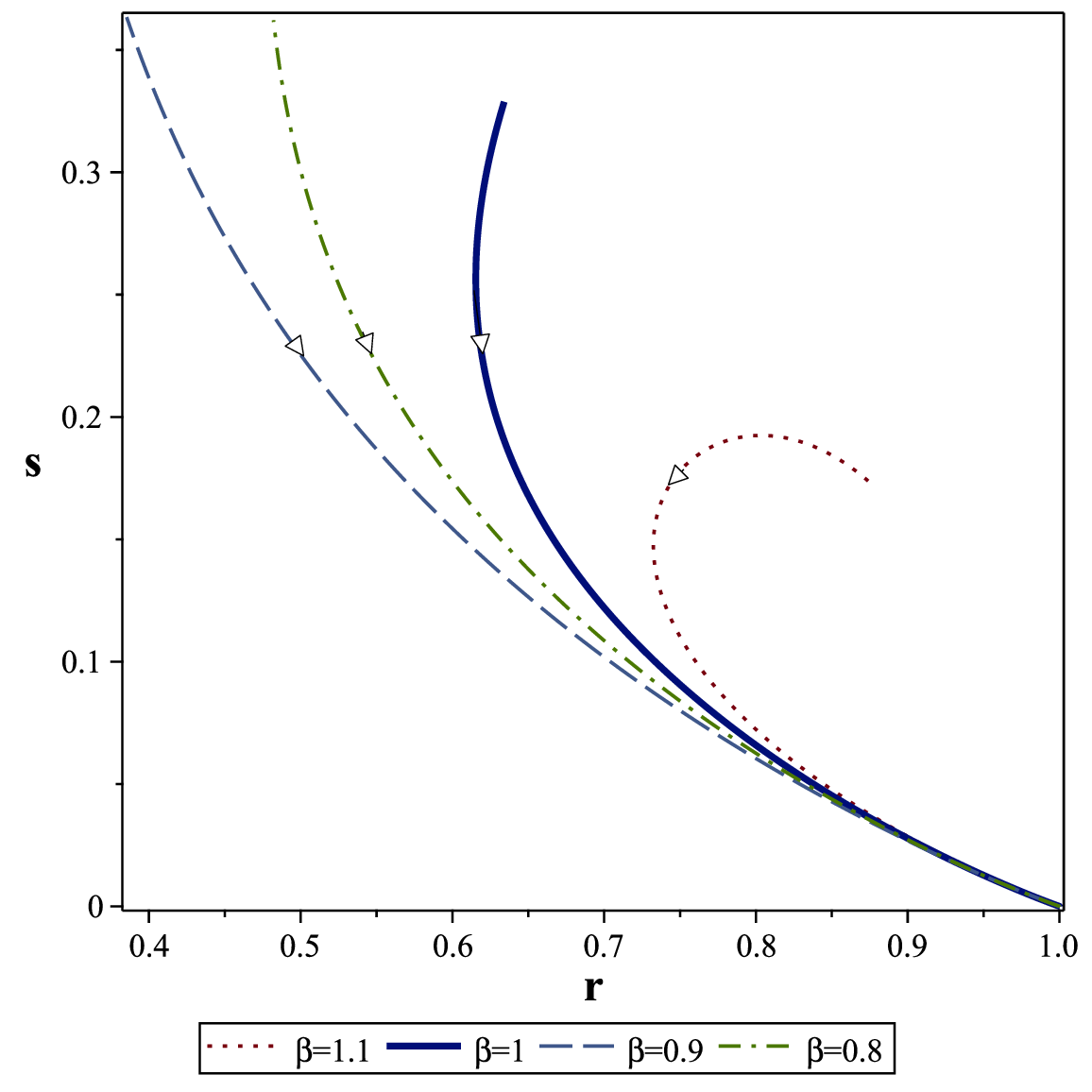}
\includegraphics[scale=0.25]{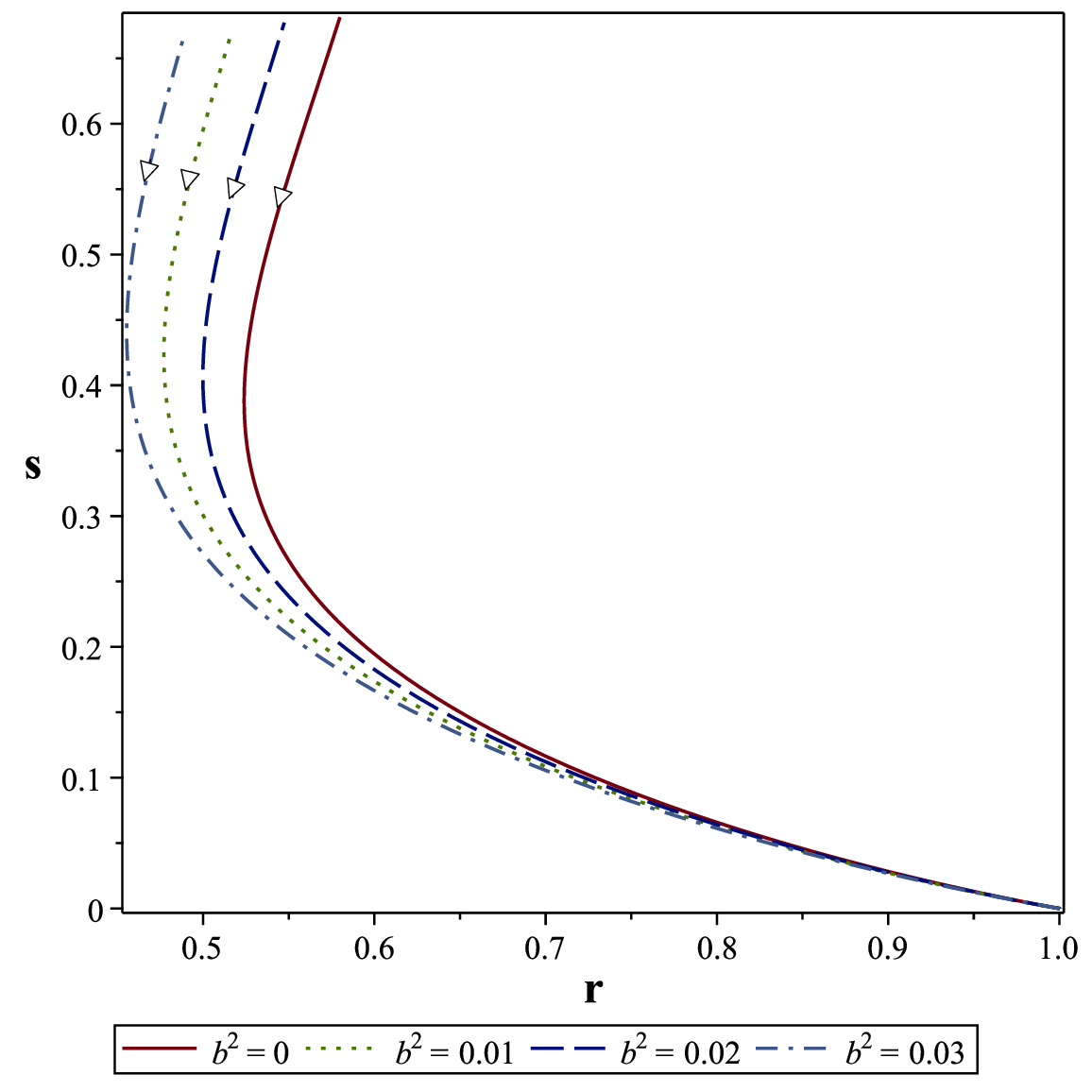}
\vspace{2mm}
\caption{In this figure we demonstrated the $\{r,s\}$ evolution. In upper panel we set $b^2=0.02$ and in the lower part we take $\beta=0.9$                                                                                                                                                                                                                                                                                                                                                                                                                                                                                                                                                                                                                                                                                                                                                                                                                                                                                                                                                                                                                                                                                                                                                                                                                                                                                                                                                                                                                                                                                                                                                                                                                                                                                                                                                                                                                                                                                                                                                                                                                                                                                                                                                                                                                                                                                                                                                                                                                                                                                                                                                                                                                                                                                                                                                                                                                                                                                                                                                                                                                                                                                                                      .} \label{tstsvrf}
\end{figure}
Next we explore the statefinder analysis for this case. To this
end we plotted \textit{r} and \textit{s} against redshift in
Figs.\ref{tstrf}-\ref{tstsvrf}. According to these figures we find
that the model can catch the point ${1,0}$ in the future while at
present time the model have a significant distance from the
$\Lambda-CDM$. Looking at $r-z$ and $s-z$ diagram, one finds that
evolutionary behavior of the model is evidently sensitive to
$\beta$($b^2$) parameter specially at past epochs while at present
and future era this sensitivity fades and all models with
different values of $\beta$ ($b^2$) approaches a same pattern.

\section{Growth of Density Perturbations in Barrow and Tsallis cosmology}
Evolution of density perturbations is of special interest in any
modified cosmology and the issue deserves a full covariant study.
Since covariant perturbation theory is beyond the scope of this
work, we just present a semi-Newotonian approach here. On this way
we follow the Jeans analysis in a two component(GDE and CDM)
universe\cite{Mukhanov_2005}. Here, we take the governing equation
of perturbation in a flat FRW universe as\cite{Mukhanov_2005}
\begin{equation}
\ddot{\delta}+2H\dot{\delta}+\left(\frac{v_s^2k^2}{a^2}-4\pi G\rho\right)\delta=0,
\end{equation}
where $\delta=\frac{\delta \rho}{\rho}$ is density contrast,
$v_s^2=\frac{\partial P}{\partial \rho}$(where P and $\rho$ are cosmic fluids pressure and energy density respectively) and $k$ denotes the comoving wave number. Here, we consider evolution of perturbations in large scales and ignore $k^2$ term(which belongs to kinetic pressure) in above equation. Beside, we need to use the first Friedmann equation in Barrow or Tsallis cosmology to disclose the effect of GDE and non-extensive nature of the model with respect to standard cosmology. To make the system of equations consistent, we also bring the evolution equation of the GDE component. We recast all of these equations versus the redshift parameter $z$. The resulting equations in Barrow cosmology read
\begin{eqnarray}
&H^2&(1+z)^2\delta^{\prime\prime}+((1+z)\frac{H^{\prime}}{H}-1)H^2(1+z)\delta^{\prime}\nonumber\\&-&\frac{3}{2}H^{2-\Delta}=0, \\
&H^{\prime}&=\frac{3H}{(1+z)(2-\Delta)}(1+\Omega_{D}w_{D}),\\
&\Omega^{\prime}&=-\frac{(1-\Delta)}{(1+z)(\Omega_D-2+\Delta)}3\Omega_D(b^2+\Omega_D-1),
\end{eqnarray}
where prime denotes derivatives with respect to $z$ and $w_D$ could be replaced from corresponding Eqs.(\ref{bwdi},\ref{wsint}). Unfortunately an analytic solution for this system of equations is not available and we ought to try a numerical task in this case. On this way we take $\Omega_D(0)=0.69, H(0)=68\frac{km}{Mpc s}, b^2=0.01$ and $\delta(100)=0.0001$. A same process is followed for the Tsallis case and the results are depicted against redshift in Fig.(\ref{delvsz}). The figure reveals that the density contrast $\delta$ is sensitive to the non-extensive parameter $\Delta$ and $\beta$. The figure reveals that in both TGDE and BGDE the structure formation has a slower rate with respect to the GDE in standard cosmology. Thus we can conclude that the non-extensivity effect can slow the structure formation process.
\begin{figure}
\centering
\includegraphics[scale=0.35]{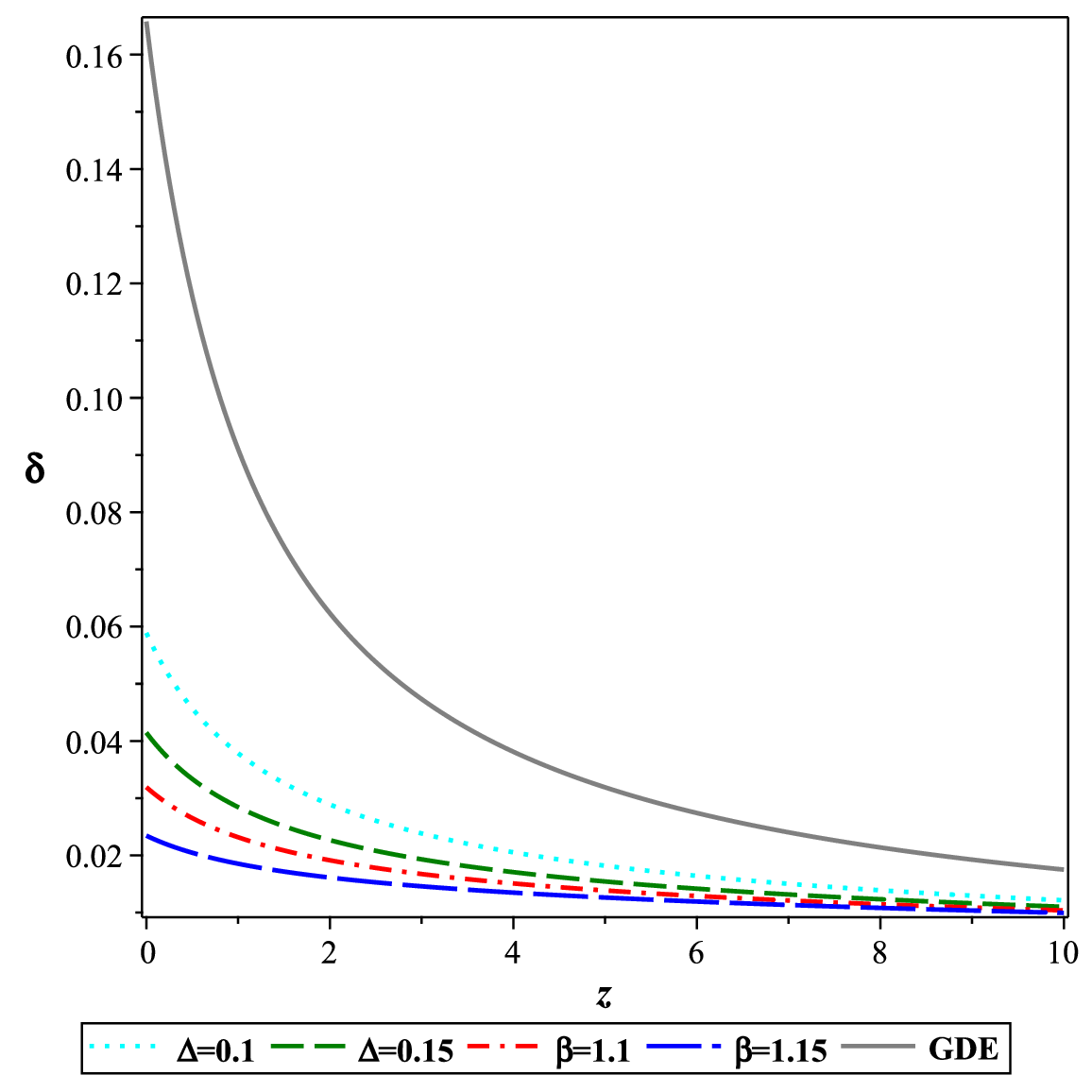}
\vspace{2mm}
\caption{Evolution of the density contrast($\delta$) is depicted against redshift for both BGDE and TGDE. In this figure we set $b^2=0.01$ for both models and also assumed $\delta(100)=0.0001$.} \label{delvsz}
\end{figure}

\section{Closing remarks}
Based on the correspondence between the laws of thermodynamics and
the gravitational field equations and using the modified
Barrow/Tsallis entropy expression, we have explored the
cosmological consequences of the modified Friedmann equations for
a universe filled with pressureless matter and DE. Here, we have
considered the GDE which has been widely investigated in the
literature. At first, we performed a dynamical system analysis to
see if the model can present a correct chronological evolution and
found that the TGDE and BGDE can exhibit such a result for
suitable range of free parameters($0<\Delta<1$ for BGDE and
$\beta<3/2$ for TGDE). One important achievement of this analysis
with respect to the GDE in standard cosmology, is appearance of a
radiation dominated phase of expansion for the choice $Q=3b^2
H(\rho_D+\rho_m)$ of interaction term. One should note that this
phase of expansion is not available for GDE in standard
cosmology\cite{Golchin:2016yci}. It is also worthy to mention that
higher limit Tsallis parameter($\beta<3/2$), is a result of
seeking an unstable phase of matter dominated in TGDE. Next, we
have explored the impacts of Barrow/Tsallis parameters on the
evolutionary description of the cosmological parameters through a
numerical task. A close look at $w_D$, $q$ and $\Omega_D$ reveals
that for both BGDE and TGDE models the transition from a
deceleration to an accelerated phase can be achieved during the
history of the universe. In Barrow/ Tsallis cosmology, larger
values of $\Delta (\beta)$ result in a delay in the time of
transition($z_{tr}$). We observed that in both BGDE and TGDE, in
the non-interacting regime, the models lie in the quintessence
region while adding a chance of interaction lead the models to
cross the phantom line. We also considered the models in presence
of an interaction term between DM and the GDE. We observed that in
both models increasing the $b^2$ leads to and earlier transition
to the accelerated expansion. The squared sound speed analysis
reveals that in BGDE(TGDE), the model is unstable against
perturbations and changing the non-extensive parameter
$\Delta(\beta)$ can not survive the model from instability issue.
We see a same result for increasing $b^2$ in both models. It is
worthy to mention that in the non-interacting case the models of
interest tend to $v^2_{s}=0$ which is the border line of stability
at far future.

We have also considered the so-called statefinder parameters
$\{r,s\}$ which are based on higher derivatives of the scale
factor and is a suitable tool in discriminating DE models
specially with respect to $\Lambda CDM$. In both underlying models
we found out that the models are significantly distinct from
$\Lambda CDM$ at high redshift epochs. Besides, comparing the
figures, $r-z$ and $s-z$, of both models reveal that these models
show a significantly different patterns of evolution. Moreover, we
saw that both BGDE and TGDE will catch the point ${r=1,s=0}$.

We have also explored the impacts of the non-extensive
parameters on the density perturbations through a Jeans analysis
and found the pattern of density contrast($\delta$) evolution in
TGDE/BGDE which differs from that of GDE in standard cosmology. We
observed that the non-extensivity make a dilation in the structure
formation process.
\newpage
\bibliography{DARK-D-24-00016-R3}

\begin{thebibliography}{10}
\expandafter\ifx\csname url\endcsname\relax
  \def\url#1{\texttt{#1}}\fi
\expandafter\ifx\csname urlprefix\endcsname\relax\def\urlprefix{URL }\fi
\expandafter\ifx\csname href\endcsname\relax
  \def\href#1#2{#2} \def\path#1{#1}\fi

\bibitem{riess982}
A.~G. {Riess}, et~al., {Observational evidence from supernovae for an
  accelerating universe and a cosmological constant}, Astron. J. 116 (1998)
  1009--1038.
\newblock \href {http://arxiv.org/abs/astro-ph/9805201}
  {\path{arXiv:astro-ph/9805201}}, \href {https://doi.org/10.1086/300499}
  {\path{doi:10.1086/300499}}.

\bibitem{riess98}
A.~G. {Riess}, A.~V. {Filippenko}, P.~Challis, A.~Clocchiatti, A.~Diercks,
  P.~M. Garnavich, R.~L. Gilliland, C.~J. Hogan, S.~Jha, R.~P. Kirshner,
  et~al., "observational evidence from supernovae for an accelerating universe
  and a cosmological constant", The Astronomical Journal 116~(3) (1998) 1009.

\bibitem{SupernovaCosmologyProject:1998vns}
S.~Perlmutter, et~al., {Measurements of $\Omega$ and $\Lambda$ from 42 high
  redshift supernovae}, Astrophys. J. 517 (1999) 565--586.
\newblock \href {http://arxiv.org/abs/astro-ph/9812133}
  {\path{arXiv:astro-ph/9812133}}, \href {https://doi.org/10.1086/307221}
  {\path{doi:10.1086/307221}}.

\bibitem{WMAP:2003elm}
D.~N. Spergel, et~al., {First year Wilkinson Microwave Anisotropy Probe (WMAP)
  observations: Determination of cosmological parameters}, Astrophys. J. Suppl.
  148 (2003) 175--194.
\newblock \href {http://arxiv.org/abs/astro-ph/0302209}
  {\path{arXiv:astro-ph/0302209}}, \href {https://doi.org/10.1086/377226}
  {\path{doi:10.1086/377226}}.

\bibitem{Capozziello:2002rd}
S.~Capozziello, {Curvature quintessence}, Int. J. Mod. Phys. D 11 (2002)
  483--492.
\newblock \href {http://arxiv.org/abs/gr-qc/0201033}
  {\path{arXiv:gr-qc/0201033}}, \href
  {https://doi.org/10.1142/S0218271802002025}
  {\path{doi:10.1142/S0218271802002025}}.

\bibitem{Sobouti:2006rd}
Y.~Sobouti, {An f(r) gravitation instead of dark matter}, Astron. Astrophys.
  464 (2007) 921, [Erratum: Astron.Astrophys. 472, 833 (2007)].
\newblock \href {http://arxiv.org/abs/0704.3345} {\path{arXiv:0704.3345}},
  \href {https://doi.org/10.1051/0004-6361:20077452}
  {\path{doi:10.1051/0004-6361:20077452}}.

\bibitem{Sotiriou:2008rp}
T.~P. Sotiriou, V.~Faraoni, {f(R) Theories Of Gravity}, Rev. Mod. Phys. 82
  (2010) 451--497.
\newblock \href {http://arxiv.org/abs/0805.1726} {\path{arXiv:0805.1726}},
  \href {https://doi.org/10.1103/RevModPhys.82.451}
  {\path{doi:10.1103/RevModPhys.82.451}}.

\bibitem{Boehmer:2007kx}
C.~G. Boehmer, T.~Harko, F.~S.~N. Lobo, {Dark matter as a geometric effect in
  f(R) gravity}, Astropart. Phys. 29 (2008) 386--392.
\newblock \href {http://arxiv.org/abs/0709.0046} {\path{arXiv:0709.0046}},
  \href {https://doi.org/10.1016/j.astropartphys.2008.04.003}
  {\path{doi:10.1016/j.astropartphys.2008.04.003}}.

\bibitem{Cognola:2007zu}
G.~Cognola, E.~Elizalde, S.~Nojiri, S.~D. Odintsov, L.~Sebastiani, S.~Zerbini,
  {A Class of viable modified f(R) gravities describing inflation and the onset
  of accelerated expansion}, Phys. Rev. D 77 (2008) 046009.
\newblock \href {http://arxiv.org/abs/0712.4017} {\path{arXiv:0712.4017}},
  \href {https://doi.org/10.1103/PhysRevD.77.046009}
  {\path{doi:10.1103/PhysRevD.77.046009}}.

\bibitem{Nojiri:2010wj}
S.~Nojiri, S.~D. Odintsov, {Unified cosmic history in modified gravity: from
  F(R) theory to Lorentz non-invariant models}, Phys. Rept. 505 (2011) 59--144.
\newblock \href {http://arxiv.org/abs/1011.0544} {\path{arXiv:1011.0544}},
  \href {https://doi.org/10.1016/j.physrep.2011.04.001}
  {\path{doi:10.1016/j.physrep.2011.04.001}}.

\bibitem{Capozziello:2011et}
S.~Capozziello, M.~De~Laurentis, {Extended Theories of Gravity}, Phys. Rept.
  509 (2011) 167--321.
\newblock \href {http://arxiv.org/abs/1108.6266} {\path{arXiv:1108.6266}},
  \href {https://doi.org/10.1016/j.physrep.2011.09.003}
  {\path{doi:10.1016/j.physrep.2011.09.003}}.

\bibitem{Odintsov:2019evb}
S.~D. Odintsov, V.~K. Oikonomou, {Unification of Inflation with Dark Energy in
  $f(R)$ Gravity and Axion Dark Matter}, Phys. Rev. D 99~(10) (2019) 104070.
\newblock \href {http://arxiv.org/abs/1905.03496} {\path{arXiv:1905.03496}},
  \href {https://doi.org/10.1103/PhysRevD.99.104070}
  {\path{doi:10.1103/PhysRevD.99.104070}}.

\bibitem{Odintsov:2019mlf}
S.~D. Odintsov, V.~K. Oikonomou, {$f(R)$ Gravity Inflation with
  String-Corrected Axion Dark Matter}, Phys. Rev. D 99~(6) (2019) 064049.
\newblock \href {http://arxiv.org/abs/1901.05363} {\path{arXiv:1901.05363}},
  \href {https://doi.org/10.1103/PhysRevD.99.064049}
  {\path{doi:10.1103/PhysRevD.99.064049}}.

\bibitem{Lanczos1932ElektromagnetismusAN}
C.~Lanczos,
  \href{https://api.semanticscholar.org/CorpusID:124628257}{Elektromagnetismus
  als nat{\"u}rliche eigenschaft der riemannschen geometrie}, Zeitschrift
  f{\"u}r Physik 73 (1932) 147--168.
\newline\urlprefix\url{https://api.semanticscholar.org/CorpusID:124628257}

\bibitem{Lanczos1938ARP}
C.~Lanczos, \href{https://api.semanticscholar.org/CorpusID:124477207}{A
  remarkable property of the riemann-christoffel tensor in four dimensions},
  Annals of Mathematics 39 (1938) 842--850.
\newline\urlprefix\url{https://api.semanticscholar.org/CorpusID:124477207}

\bibitem{Dehghani:2004cf}
M.~H. Dehghani, {Accelerated expansion of the Universe in Gauss-Bonnet
  gravity}, Phys. Rev. D 70 (2004) 064009.
\newblock \href {http://arxiv.org/abs/hep-th/0404118}
  {\path{arXiv:hep-th/0404118}}, \href
  {https://doi.org/10.1103/PhysRevD.70.064009}
  {\path{doi:10.1103/PhysRevD.70.064009}}.

\bibitem{Lovelock1969DivergencefreeTC}
D.~Lovelock,
  \href{https://api.semanticscholar.org/CorpusID:189832116}{Divergence-free
  tensorial concomitants}, aequationes mathematicae 4 (1969) 127--138.
\newline\urlprefix\url{https://api.semanticscholar.org/CorpusID:189832116}

\bibitem{Jacobson:1995ab}
T.~Jacobson, {Thermodynamics of space-time: The Einstein equation of state},
  Phys. Rev. Lett. 75 (1995) 1260--1263.
\newblock \href {http://arxiv.org/abs/gr-qc/9504004}
  {\path{arXiv:gr-qc/9504004}}, \href
  {https://doi.org/10.1103/PhysRevLett.75.1260}
  {\path{doi:10.1103/PhysRevLett.75.1260}}.

\bibitem{Verlinde:2010hp}
E.~P. Verlinde, {On the Origin of Gravity and the Laws of Newton}, JHEP 04
  (2011) 029.
\newblock \href {http://arxiv.org/abs/1001.0785} {\path{arXiv:1001.0785}},
  \href {https://doi.org/10.1007/JHEP04(2011)029}
  {\path{doi:10.1007/JHEP04(2011)029}}.

\bibitem{Cai:2010hk}
R.-G. Cai, L.-M. Cao, N.~Ohta, {Friedmann Equations from Entropic Force}, Phys.
  Rev. D 81 (2010) 061501.
\newblock \href {http://arxiv.org/abs/1001.3470} {\path{arXiv:1001.3470}},
  \href {https://doi.org/10.1103/PhysRevD.81.061501}
  {\path{doi:10.1103/PhysRevD.81.061501}}.

\bibitem{Cai:2010sz}
R.-G. Cai, L.-M. Cao, N.~Ohta, {Notes on Entropy Force in General Spherically
  Symmetric Spacetimes}, Phys. Rev. D 81 (2010) 084012.
\newblock \href {http://arxiv.org/abs/1002.1136} {\path{arXiv:1002.1136}},
  \href {https://doi.org/10.1103/PhysRevD.81.084012}
  {\path{doi:10.1103/PhysRevD.81.084012}}.

\bibitem{Banerjee:2010yd}
R.~Banerjee, B.~R. Majhi, {Statistical Origin of Gravity}, Phys. Rev. D 81
  (2010) 124006.
\newblock \href {http://arxiv.org/abs/1003.2312} {\path{arXiv:1003.2312}},
  \href {https://doi.org/10.1103/PhysRevD.81.124006}
  {\path{doi:10.1103/PhysRevD.81.124006}}.

\bibitem{Liu:2010na}
Y.-X. Liu, Y.-Q. Wang, S.-W. Wei, {A Note on Temperature and Energy of
  4-dimensional Black Holes from Entropic Force}, Class. Quant. Grav. 27 (2010)
  185002.
\newblock \href {http://arxiv.org/abs/1002.1062} {\path{arXiv:1002.1062}},
  \href {https://doi.org/10.1088/0264-9381/27/18/185002}
  {\path{doi:10.1088/0264-9381/27/18/185002}}.

\bibitem{Sheykhi:2010wm}
A.~Sheykhi, {Entropic Corrections to Friedmann Equations}, Phys. Rev. D 81
  (2010) 104011.
\newblock \href {http://arxiv.org/abs/1004.0627} {\path{arXiv:1004.0627}},
  \href {https://doi.org/10.1103/PhysRevD.81.104011}
  {\path{doi:10.1103/PhysRevD.81.104011}}.

\bibitem{Modesto:2010rm}
L.~Modesto, A.~Randono, {Entropic Corrections to Newton's Law} (3 2010).
\newblock \href {http://arxiv.org/abs/1003.1998} {\path{arXiv:1003.1998}}.

\bibitem{Cai:2010zw}
Y.-F. Cai, J.~Liu, H.~Li, {Entropic cosmology: a unified model of inflation and
  late-time acceleration}, Phys. Lett. B 690 (2010) 213--219.
\newblock \href {http://arxiv.org/abs/1003.4526} {\path{arXiv:1003.4526}},
  \href {https://doi.org/10.1016/j.physletb.2010.05.033}
  {\path{doi:10.1016/j.physletb.2010.05.033}}.

\bibitem{Hendi:2010xr}
S.~H. Hendi, A.~Sheykhi, {Entropic Corrections to Einstein Equations}, Phys.
  Rev. D 83 (2011) 084012.
\newblock \href {http://arxiv.org/abs/1012.0381} {\path{arXiv:1012.0381}},
  \href {https://doi.org/10.1103/PhysRevD.83.084012}
  {\path{doi:10.1103/PhysRevD.83.084012}}.

\bibitem{Sheykhi:2010yq}
A.~Sheykhi, S.~H. Hendi, {Power-Law Entropic Corrections to Newton's Law and
  Friedmann Equations}, Phys. Rev. D 84 (2011) 044023.
\newblock \href {http://arxiv.org/abs/1011.0676} {\path{arXiv:1011.0676}},
  \href {https://doi.org/10.1103/PhysRevD.84.044023}
  {\path{doi:10.1103/PhysRevD.84.044023}}.

\bibitem{Tsallis:1987eu}
C.~Tsallis, {Possible Generalization of Boltzmann-Gibbs Statistics}, J.
  Statist. Phys. 52 (1988) 479--487.
\newblock \href {https://doi.org/10.1007/BF01016429}
  {\path{doi:10.1007/BF01016429}}.

\bibitem{Lyra:1997ggy}
M.~L. Lyra, C.~Tsallis, {Nonextensivity and Multifractality in Low-Dimensional
  Dissipative Systems}, Phys. Rev. Lett. 80 (1998) 53--56.
\newblock \href {http://arxiv.org/abs/cond-mat/9709226}
  {\path{arXiv:cond-mat/9709226}}, \href
  {https://doi.org/10.1103/PhysRevLett.80.53}
  {\path{doi:10.1103/PhysRevLett.80.53}}.

\bibitem{Wilk:1999dr}
G.~Wilk, Z.~Wlodarczyk, {On the interpretation of nonextensive parameter q in
  Tsallis statistics and Levy distributions}, Phys. Rev. Lett. 84 (2000) 2770.
\newblock \href {http://arxiv.org/abs/hep-ph/9908459}
  {\path{arXiv:hep-ph/9908459}}, \href
  {https://doi.org/10.1103/PhysRevLett.84.2770}
  {\path{doi:10.1103/PhysRevLett.84.2770}}.

\bibitem{Tsallis:2012js}
C.~Tsallis, L.~J.~L. Cirto, {Black hole thermodynamical entropy}, Eur. Phys. J.
  C 73 (2013) 2487.
\newblock \href {http://arxiv.org/abs/1202.2154} {\path{arXiv:1202.2154}},
  \href {https://doi.org/10.1140/epjc/s10052-013-2487-6}
  {\path{doi:10.1140/epjc/s10052-013-2487-6}}.

\bibitem{Tavayef:2018xwx}
M.~Tavayef, A.~Sheykhi, K.~Bamba, H.~Moradpour, {Tsallis Holographic Dark
  Energy}, Phys. Lett. B 781 (2018) 195--200.
\newblock \href {http://arxiv.org/abs/1804.02983} {\path{arXiv:1804.02983}},
  \href {https://doi.org/10.1016/j.physletb.2018.04.001}
  {\path{doi:10.1016/j.physletb.2018.04.001}}.

\bibitem{Zadeh:2018poj}
M.~A. Zadeh, A.~Sheykhi, H.~Moradpour, K.~Bamba, {Note on Tsallis holographic
  dark energy}, Eur. Phys. J. C 78~(11) (2018) 940.
\newblock \href {http://arxiv.org/abs/1806.07285} {\path{arXiv:1806.07285}},
  \href {https://doi.org/10.1140/epjc/s10052-018-6427-3}
  {\path{doi:10.1140/epjc/s10052-018-6427-3}}.

\bibitem{AbdollahiZadeh:2019lsx}
M.~Abdollahi~Zadeh, A.~Sheykhi, H.~Moradpour, {Thermal stability of Tsallis
  holographic dark energy in nonflat universe}, Gen. Rel. Grav. 51~(1) (2019)
  12.
\newblock \href {https://doi.org/10.1007/s10714-018-2497-7}
  {\path{doi:10.1007/s10714-018-2497-7}}.

\bibitem{Pandey:2021fvr}
B.~D. Pandey, S.~K. P, Pankaj, U.~K. Sharma, {New Tsallis holographic dark
  energy}, Eur. Phys. J. C 82~(3) (2022) 233.
\newblock \href {http://arxiv.org/abs/2110.13628} {\path{arXiv:2110.13628}},
  \href {https://doi.org/10.1140/epjc/s10052-022-10171-w}
  {\path{doi:10.1140/epjc/s10052-022-10171-w}}.

\bibitem{Huang:2019hex}
Q.~Huang, H.~Huang, J.~Chen, L.~Zhang, F.~Tu, {Stability analysis of a Tsallis
  holographic dark energy model}, Class. Quant. Grav. 36~(17) (2019) 175001.
\newblock \href {http://arxiv.org/abs/2201.12504} {\path{arXiv:2201.12504}},
  \href {https://doi.org/10.1088/1361-6382/ab3504}
  {\path{doi:10.1088/1361-6382/ab3504}}.

\bibitem{Sheykhi:2018dpn}
A.~Sheykhi, {Modified Friedmann Equations from Tsallis Entropy}, Phys. Lett. B
  785 (2018) 118--126.
\newblock \href {http://arxiv.org/abs/1806.03996} {\path{arXiv:1806.03996}},
  \href {https://doi.org/10.1016/j.physletb.2018.08.036}
  {\path{doi:10.1016/j.physletb.2018.08.036}}.

\bibitem{Asghari:2021lzu}
M.~Asghari, A.~Sheykhi, {Observational constraints on Tsallis modified
  gravity}, Mon. Not. Roy. Astron. Soc. 508~(2) (2021) 2855--2861.
\newblock \href {http://arxiv.org/abs/2106.15551} {\path{arXiv:2106.15551}},
  \href {https://doi.org/10.1093/mnras/stab2671}
  {\path{doi:10.1093/mnras/stab2671}}.

\bibitem{Sheykhi:2022gzb}
A.~Sheykhi, B.~Farsi, {Growth of perturbations in Tsallis and Barrow
  cosmology}, Eur. Phys. J. C 82~(12) (2022) 1111.
\newblock \href {http://arxiv.org/abs/2205.04138} {\path{arXiv:2205.04138}},
  \href {https://doi.org/10.1140/epjc/s10052-022-11044-y}
  {\path{doi:10.1140/epjc/s10052-022-11044-y}}.

\bibitem{Basilakos:2023kvk}
S.~Basilakos, A.~Lymperis, M.~Petronikolou, E.~N. Saridakis, {Alleviating both
  $H_0$ and $\sigma_8$ tensions in Tsallis cosmology} (8 2023).
\newblock \href {http://arxiv.org/abs/2308.01200} {\path{arXiv:2308.01200}}.

\bibitem{Luciano:2022ely}
G.~G. Luciano, J.~Gine, {Baryogenesis in non-extensive Tsallis Cosmology},
  Phys. Lett. B 833 (2022) 137352.
\newblock \href {http://arxiv.org/abs/2204.02723} {\path{arXiv:2204.02723}},
  \href {https://doi.org/10.1016/j.physletb.2022.137352}
  {\path{doi:10.1016/j.physletb.2022.137352}}.

\bibitem{Barrow:2020tzx}
J.~D. Barrow, {The Area of a Rough Black Hole}, Phys. Lett. B 808 (2020)
  135643.
\newblock \href {http://arxiv.org/abs/2004.09444} {\path{arXiv:2004.09444}},
  \href {https://doi.org/10.1016/j.physletb.2020.135643}
  {\path{doi:10.1016/j.physletb.2020.135643}}.

\bibitem{Sheykhi:2021fwh}
A.~Sheykhi, {Barrow Entropy Corrections to Friedmann Equations}, Phys. Rev. D
  103~(12) (2021) 123503.
\newblock \href {http://arxiv.org/abs/2102.06550} {\path{arXiv:2102.06550}},
  \href {https://doi.org/10.1103/PhysRevD.103.123503}
  {\path{doi:10.1103/PhysRevD.103.123503}}.

\bibitem{Saridakis:2020lrg}
E.~N. Saridakis, {Modified cosmology through spacetime thermodynamics and
  Barrow horizon entropy}, JCAP 07 (2020) 031.
\newblock \href {http://arxiv.org/abs/2006.01105} {\path{arXiv:2006.01105}},
  \href {https://doi.org/10.1088/1475-7516/2020/07/031}
  {\path{doi:10.1088/1475-7516/2020/07/031}}.

\bibitem{Saridakis:2020cqq}
E.~N. Saridakis, S.~Basilakos, {The generalized second law of thermodynamics
  with Barrow entropy}, Eur. Phys. J. C 81~(7) (2021) 644.
\newblock \href {http://arxiv.org/abs/2005.08258} {\path{arXiv:2005.08258}},
  \href {https://doi.org/10.1140/epjc/s10052-021-09431-y}
  {\path{doi:10.1140/epjc/s10052-021-09431-y}}.

\bibitem{Saridakis:2020zol}
E.~N. Saridakis, {Barrow holographic dark energy}, Phys. Rev. D 102~(12) (2020)
  123525.
\newblock \href {http://arxiv.org/abs/2005.04115} {\path{arXiv:2005.04115}},
  \href {https://doi.org/10.1103/PhysRevD.102.123525}
  {\path{doi:10.1103/PhysRevD.102.123525}}.

\bibitem{Srivastava:2020cyk}
S.~Srivastava, U.~K. Sharma, {Barrow holographic dark energy with Hubble
  horizon as IR cutoff}, Int. J. Geom. Meth. Mod. Phys. 18~(01) (2021) 2150014.
\newblock \href {http://arxiv.org/abs/2010.09439} {\path{arXiv:2010.09439}},
  \href {https://doi.org/10.1142/S0219887821500146}
  {\path{doi:10.1142/S0219887821500146}}.

\bibitem{Adhikary:2021xym}
P.~Adhikary, S.~Das, S.~Basilakos, E.~N. Saridakis, {Barrow holographic dark
  energy in a nonflat universe}, Phys. Rev. D 104~(12) (2021) 123519.
\newblock \href {http://arxiv.org/abs/2104.13118} {\path{arXiv:2104.13118}},
  \href {https://doi.org/10.1103/PhysRevD.104.123519}
  {\path{doi:10.1103/PhysRevD.104.123519}}.

\bibitem{Oliveros:2022biu}
A.~Oliveros, M.~A. Sabogal, M.~A. Acero, {Barrow holographic dark energy with
  Granda\textendash{}Oliveros cutoff}, Eur. Phys. J. Plus 137~(7) (2022) 783.
\newblock \href {http://arxiv.org/abs/2203.14464} {\path{arXiv:2203.14464}},
  \href {https://doi.org/10.1140/epjp/s13360-022-02994-z}
  {\path{doi:10.1140/epjp/s13360-022-02994-z}}.

\bibitem{Anagnostopoulos:2020ctz}
F.~K. Anagnostopoulos, S.~Basilakos, E.~N. Saridakis, {Observational
  constraints on Barrow holographic dark energy}, Eur. Phys. J. C 80~(9) (2020)
  826.
\newblock \href {http://arxiv.org/abs/2005.10302} {\path{arXiv:2005.10302}},
  \href {https://doi.org/10.1140/epjc/s10052-020-8360-5}
  {\path{doi:10.1140/epjc/s10052-020-8360-5}}.

\bibitem{Dabrowski:2020atl}
M.~P. Dabrowski, V.~Salzano, {Geometrical observational bounds on a fractal
  horizon holographic dark energy}, Phys. Rev. D 102~(6) (2020) 064047.
\newblock \href {http://arxiv.org/abs/2009.08306} {\path{arXiv:2009.08306}},
  \href {https://doi.org/10.1103/PhysRevD.102.064047}
  {\path{doi:10.1103/PhysRevD.102.064047}}.

\bibitem{veneziano1979u}
G.~Veneziano, U (1) without instantons, Nuclear Physics B 159~(1-2) (1979)
  213--224.

\bibitem{Witten:1979vv}
E.~Witten, {Current Algebra Theorems for the U(1) Goldstone Boson}, Nucl. Phys.
  B 156 (1979) 269--283.
\newblock \href {https://doi.org/10.1016/0550-3213(79)90031-2}
  {\path{doi:10.1016/0550-3213(79)90031-2}}.

\bibitem{Rosenzweig:1979ay}
C.~Rosenzweig, J.~Schechter, C.~G. Trahern, {Is the Effective Lagrangian for
  QCD a Sigma Model?}, Phys. Rev. D 21 (1980) 3388.
\newblock \href {https://doi.org/10.1103/PhysRevD.21.3388}
  {\path{doi:10.1103/PhysRevD.21.3388}}.

\bibitem{Nath:1979ik}
P.~Nath, R.~L. Arnowitt, {The U(1) Problem: Current Algebra and the Theta
  Vacuum}, Phys. Rev. D 23 (1981) 473.
\newblock \href {https://doi.org/10.1103/PhysRevD.23.473}
  {\path{doi:10.1103/PhysRevD.23.473}}.

\bibitem{Kawarabayashi:1980dp}
K.~Kawarabayashi, N.~Ohta, {The Problem of $\eta$ in the Large $N$ Limit:
  Effective Lagrangian Approach}, Nucl. Phys. B 175 (1980) 477--492.
\newblock \href {https://doi.org/10.1016/0550-3213(80)90024-3}
  {\path{doi:10.1016/0550-3213(80)90024-3}}.

\bibitem{Ohta:2010in}
N.~Ohta, {Dark Energy and QCD Ghost}, Phys. Lett. B 695 (2011) 41--44.
\newblock \href {http://arxiv.org/abs/1010.1339} {\path{arXiv:1010.1339}},
  \href {https://doi.org/10.1016/j.physletb.2010.11.044}
  {\path{doi:10.1016/j.physletb.2010.11.044}}.

\bibitem{Urban:2009vy}
F.~R. Urban, A.~R. Zhitnitsky, {The cosmological constant from the QCD
  Veneziano ghost}, Phys. Lett. B 688 (2010) 9--12.
\newblock \href {http://arxiv.org/abs/0906.2162} {\path{arXiv:0906.2162}},
  \href {https://doi.org/10.1016/j.physletb.2010.03.080}
  {\path{doi:10.1016/j.physletb.2010.03.080}}.

\bibitem{Cai:2010uf}
R.-G. Cai, Z.-L. Tuo, H.-B. Zhang, Q.~Su, {Notes on Ghost Dark Energy}, Phys.
  Rev. D 84 (2011) 123501.
\newblock \href {http://arxiv.org/abs/1011.3212} {\path{arXiv:1011.3212}},
  \href {https://doi.org/10.1103/PhysRevD.84.123501}
  {\path{doi:10.1103/PhysRevD.84.123501}}.

\bibitem{Sheykhi:2011xz}
A.~Sheykhi, M.~Sadegh~Movahed, {Interacting Ghost Dark Energy in Non-Flat
  Universe}, Gen. Rel. Grav. 44 (2012) 449--465.
\newblock \href {http://arxiv.org/abs/1104.4713} {\path{arXiv:1104.4713}},
  \href {https://doi.org/10.1007/s10714-011-1286-3}
  {\path{doi:10.1007/s10714-011-1286-3}}.

\bibitem{Ebrahimi:2011js}
E.~Ebrahimi, A.~Sheykhi, {Instability of QCD ghost dark energy model}, Int. J.
  Mod. Phys. D 20 (2011) 2369--2381.
\newblock \href {http://arxiv.org/abs/1106.3504} {\path{arXiv:1106.3504}},
  \href {https://doi.org/10.1142/S021827181102041X}
  {\path{doi:10.1142/S021827181102041X}}.

\bibitem{Malekjani:2012wc}
M.~Malekjani, A.~Khodam-Mohammadi, {Statefinder diagnosis and the interacting
  ghost model of dark energy}, Astrophys. Space Sci. 343 (2013) 451--461.
\newblock \href {http://arxiv.org/abs/1202.4154} {\path{arXiv:1202.4154}},
  \href {https://doi.org/10.1007/s10509-012-1230-3}
  {\path{doi:10.1007/s10509-012-1230-3}}.

\bibitem{Liu:2020bmp}
Y.~Liu, {Interacting ghost dark energy in complex quintessence theory}, Eur.
  Phys. J. C 80~(12) (2020) 1204.
\newblock \href {http://arxiv.org/abs/2201.00658} {\path{arXiv:2201.00658}},
  \href {https://doi.org/10.1140/epjc/s10052-020-08786-y}
  {\path{doi:10.1140/epjc/s10052-020-08786-y}}.

\bibitem{Sheykhi:2022fus}
A.~Sheykhi, M.~S. Hamedan, {Holographic Dark Energy in Modified Barrow
  Cosmology}, Entropy 25~(4) (2023) 569.
\newblock \href {http://arxiv.org/abs/2211.00088} {\path{arXiv:2211.00088}},
  \href {https://doi.org/10.3390/e25040569} {\path{doi:10.3390/e25040569}}.

\bibitem{Sheykhi:2023woy}
A.~Sheykhi, S.~Ghaffari, {Note on agegraphic dark energy inspired by modified
  Barrow entropy}, Phys. Dark Univ. 41 (2023) 101241.
\newblock \href {http://arxiv.org/abs/2304.03261} {\path{arXiv:2304.03261}},
  \href {https://doi.org/10.1016/j.dark.2023.101241}
  {\path{doi:10.1016/j.dark.2023.101241}}.

\bibitem{Cai:2009ph}
R.-G. Cai, N.~Ohta, {Horizon Thermodynamics and Gravitational Field Equations
  in Horava-Lifshitz Gravity}, Phys. Rev. D 81 (2010) 084061.
\newblock \href {http://arxiv.org/abs/0910.2307} {\path{arXiv:0910.2307}},
  \href {https://doi.org/10.1103/PhysRevD.81.084061}
  {\path{doi:10.1103/PhysRevD.81.084061}}.

\bibitem{Hayward:1997jp}
S.~A. Hayward, {Unified first law of black hole dynamics and relativistic
  thermodynamics}, Class. Quant. Grav. 15 (1998) 3147--3162.
\newblock \href {http://arxiv.org/abs/gr-qc/9710089}
  {\path{arXiv:gr-qc/9710089}}, \href
  {https://doi.org/10.1088/0264-9381/15/10/017}
  {\path{doi:10.1088/0264-9381/15/10/017}}.

\bibitem{Bertolami:2007zm}
O.~Bertolami, F.~Gil~Pedro, M.~Le~Delliou, {Dark Energy-Dark Matter Interaction
  and the Violation of the Equivalence Principle from the Abell Cluster A586},
  Phys. Lett. B 654 (2007) 165--169.
\newblock \href {http://arxiv.org/abs/astro-ph/0703462}
  {\path{arXiv:astro-ph/0703462}}, \href
  {https://doi.org/10.1016/j.physletb.2007.08.046}
  {\path{doi:10.1016/j.physletb.2007.08.046}}.

\bibitem{Wetterich:1987fm}
C.~Wetterich, {Cosmology and the Fate of Dilatation Symmetry}, Nucl. Phys. B
  302 (1988) 668--696.
\newblock \href {http://arxiv.org/abs/1711.03844} {\path{arXiv:1711.03844}},
  \href {https://doi.org/10.1016/0550-3213(88)90193-9}
  {\path{doi:10.1016/0550-3213(88)90193-9}}.

\bibitem{Amendola:1999qq}
L.~Amendola, {Scaling solutions in general nonminimal coupling theories}, Phys.
  Rev. D 60 (1999) 043501.
\newblock \href {http://arxiv.org/abs/astro-ph/9904120}
  {\path{arXiv:astro-ph/9904120}}, \href
  {https://doi.org/10.1103/PhysRevD.60.043501}
  {\path{doi:10.1103/PhysRevD.60.043501}}.

\bibitem{Amendola:2000uh}
L.~Amendola, D.~Tocchini-Valentini, {Stationary dark energy: The Present
  universe as a global attractor}, Phys. Rev. D 64 (2001) 043509.
\newblock \href {http://arxiv.org/abs/astro-ph/0011243}
  {\path{arXiv:astro-ph/0011243}}, \href
  {https://doi.org/10.1103/PhysRevD.64.043509}
  {\path{doi:10.1103/PhysRevD.64.043509}}.

\bibitem{Zimdahl:2001ar}
W.~Zimdahl, D.~Pavon, {Interacting quintessence}, Phys. Lett. B 521 (2001)
  133--138.
\newblock \href {http://arxiv.org/abs/astro-ph/0105479}
  {\path{arXiv:astro-ph/0105479}}, \href
  {https://doi.org/10.1016/S0370-2693(01)01174-1}
  {\path{doi:10.1016/S0370-2693(01)01174-1}}.

\bibitem{Zimdahl:2002zb}
W.~Zimdahl, D.~Pav\'on, {Scaling cosmology}, Gen. Rel. Grav. 35 (2003)
  413--422.
\newblock \href {http://arxiv.org/abs/astro-ph/0210484}
  {\path{arXiv:astro-ph/0210484}}, \href
  {https://doi.org/10.1023/A:1022369800053}
  {\path{doi:10.1023/A:1022369800053}}.

\bibitem{Chimento:2003iea}
L.~P. Chimento, A.~S. Jakubi, D.~Pavon, W.~Zimdahl, {Interacting quintessence
  solution to the coincidence problem}, Phys. Rev. D 67 (2003) 083513.
\newblock \href {http://arxiv.org/abs/astro-ph/0303145}
  {\path{arXiv:astro-ph/0303145}}, \href
  {https://doi.org/10.1103/PhysRevD.67.083513}
  {\path{doi:10.1103/PhysRevD.67.083513}}.

\bibitem{Wang:2005jx}
B.~Wang, Y.-g. Gong, E.~Abdalla, {Transition of the dark energy equation of
  state in an interacting holographic dark energy model}, Phys. Lett. B 624
  (2005) 141--146.
\newblock \href {http://arxiv.org/abs/hep-th/0506069}
  {\path{arXiv:hep-th/0506069}}, \href
  {https://doi.org/10.1016/j.physletb.2005.08.008}
  {\path{doi:10.1016/j.physletb.2005.08.008}}.

\bibitem{Wang:2005ph}
B.~Wang, C.-Y. Lin, E.~Abdalla, {Constraints on the interacting holographic
  dark energy model}, Phys. Lett. B 637 (2006) 357--361.
\newblock \href {http://arxiv.org/abs/hep-th/0509107}
  {\path{arXiv:hep-th/0509107}}, \href
  {https://doi.org/10.1016/j.physletb.2006.04.009}
  {\path{doi:10.1016/j.physletb.2006.04.009}}.

\bibitem{Mangano:2002gg}
G.~Mangano, G.~Miele, V.~Pettorino, {Coupled quintessence and the coincidence
  problem}, Mod. Phys. Lett. A 18 (2003) 831--842.
\newblock \href {http://arxiv.org/abs/astro-ph/0212518}
  {\path{arXiv:astro-ph/0212518}}, \href
  {https://doi.org/10.1142/S0217732303009940}
  {\path{doi:10.1142/S0217732303009940}}.

\bibitem{Arevalo:2011hh}
F.~Arevalo, A.~P.~R. Bacalhau, W.~Zimdahl, {Cosmological dynamics with
  non-linear interactions}, Class. Quant. Grav. 29 (2012) 235001.
\newblock \href {http://arxiv.org/abs/1112.5095} {\path{arXiv:1112.5095}},
  \href {https://doi.org/10.1088/0264-9381/29/23/235001}
  {\path{doi:10.1088/0264-9381/29/23/235001}}.

\bibitem{Baldi:2010vv}
M.~Baldi, {Time dependent couplings in the dark sector: from background
  evolution to nonlinear structure formation}, Mon. Not. Roy. Astron. Soc. 411
  (2011) 1077.
\newblock \href {http://arxiv.org/abs/1005.2188} {\path{arXiv:1005.2188}},
  \href {https://doi.org/10.1111/j.1365-2966.2010.17758.x}
  {\path{doi:10.1111/j.1365-2966.2010.17758.x}}.

\bibitem{Amendola:1999er}
L.~Amendola, {Coupled quintessence}, Phys. Rev. D 62 (2000) 043511.
\newblock \href {http://arxiv.org/abs/astro-ph/9908023}
  {\path{arXiv:astro-ph/9908023}}, \href
  {https://doi.org/10.1103/PhysRevD.62.043511}
  {\path{doi:10.1103/PhysRevD.62.043511}}.

\bibitem{xu2012phase}
C.~Xu, E.~N. Saridakis, G.~Leon, Phase-space analysis of teleparallel dark
  energy, Journal of Cosmology and Astroparticle Physics 2012~(07) (2012) 005.

\bibitem{Landim:2015uda}
R.~C.~G. Landim, {Coupled dark energy: a dynamical analysis with complex scalar
  field}, Eur. Phys. J. C 76~(1) (2016) 31.
\newblock \href {http://arxiv.org/abs/1507.00902} {\path{arXiv:1507.00902}},
  \href {https://doi.org/10.1140/epjc/s10052-016-3894-2}
  {\path{doi:10.1140/epjc/s10052-016-3894-2}}.

\bibitem{Landim:2015poa}
R.~C.~G. Landim, {Coupled tachyonic dark energy: a dynamical analysis}, Int. J.
  Mod. Phys. D 24~(11) (2015) 1550085.
\newblock \href {http://arxiv.org/abs/1505.03243} {\path{arXiv:1505.03243}},
  \href {https://doi.org/10.1142/S0218271815500856}
  {\path{doi:10.1142/S0218271815500856}}.

\bibitem{Leon:2021wyx}
G.~Leon, J.~Maga\~na, A.~Hern\'andez-Almada, M.~A. Garc\'\i{}a-Aspeitia,
  T.~Verdugo, V.~Motta, {Barrow Entropy Cosmology: an observational approach
  with a hint of stability analysis}, JCAP 12~(12) (2021) 032.
\newblock \href {http://arxiv.org/abs/2108.10998} {\path{arXiv:2108.10998}},
  \href {https://doi.org/10.1088/1475-7516/2021/12/032}
  {\path{doi:10.1088/1475-7516/2021/12/032}}.

\bibitem{Golchin:2016yci}
H.~Golchin, S.~Jamali, E.~Ebrahimi, {Interacting Dark Energy: Dynamical System
  Analysis}, Int. J. Mod. Phys. D 26~(09) (2017) 1750098.
\newblock \href {http://arxiv.org/abs/1605.05068} {\path{arXiv:1605.05068}},
  \href {https://doi.org/10.1142/S0218271817500985}
  {\path{doi:10.1142/S0218271817500985}}.

\bibitem{Planck:2018vyg}
N.~Aghanim, et~al., {Planck 2018 results. VI. Cosmological parameters}, Astron.
  Astrophys. 641 (2020) A6, [Erratum: Astron.Astrophys. 652, C4 (2021)].
\newblock \href {http://arxiv.org/abs/1807.06209} {\path{arXiv:1807.06209}},
  \href {https://doi.org/10.1051/0004-6361/201833910}
  {\path{doi:10.1051/0004-6361/201833910}}.

\bibitem{Sahni:2002fz}
V.~Sahni, T.~D. Saini, A.~A. Starobinsky, U.~Alam, {Statefinder: A New
  geometrical diagnostic of dark energy}, JETP Lett. 77 (2003) 201--206.
\newblock \href {http://arxiv.org/abs/astro-ph/0201498}
  {\path{arXiv:astro-ph/0201498}}, \href {https://doi.org/10.1134/1.1574831}
  {\path{doi:10.1134/1.1574831}}.

\bibitem{Daly:2007dn}
R.~A. Daly, S.~G. Djorgovski, K.~A. Freeman, M.~P. Mory, C.~P. O'Dea, P.~Kharb,
  S.~Baum, {Improved Constraints on the Acceleration History of the Universe
  and the Properties of the Dark Energy}, Astrophys. J. 677 (2008) 1--11.
\newblock \href {http://arxiv.org/abs/0710.5345} {\path{arXiv:0710.5345}},
  \href {https://doi.org/10.1086/528837} {\path{doi:10.1086/528837}}.

\bibitem{WMAP:2010qai}
E.~Komatsu, et~al., {Seven-Year Wilkinson Microwave Anisotropy Probe (WMAP)
  Observations: Cosmological Interpretation}, Astrophys. J. Suppl. 192 (2011)
  18.
\newblock \href {http://arxiv.org/abs/1001.4538} {\path{arXiv:1001.4538}},
  \href {https://doi.org/10.1088/0067-0049/192/2/18}
  {\path{doi:10.1088/0067-0049/192/2/18}}.

\bibitem{Mukhanov_2005}
V.~Mukhanov, Physical Foundations of Cosmology, Cambridge University Press,
  2005.

\end{thebibliography}

\end{document}